\documentclass[prodmode]{acmsmall} 

\setcopyright{rightsretained} 

\usepackage{tabularx}
\usepackage[utf8]{inputenc}
\usepackage{natbib}
\usepackage{indentfirst}
\usepackage{graphicx}
\usepackage{wrapfig}
\usepackage[a4paper]{geometry}
\usepackage{fullpage}
\usepackage{tabulary}
\usepackage{hyperref}

\usepackage{lscape}
\usepackage{longtable}
\usepackage{calc}
\usepackage{multirow}

\begin{document}

\markboth{K. H\"ansel et al.}{Wearable Computing for Health: Exploring the Relationship between Data and Human Behaviour}

\title{Wearable Computing for Health and Fitness: Exploring the Relationship between Data and Human Behaviour}
\author{KATRIN H\"ANSEL
\affil{Queen Mary University of London}
NATALIE WILDE 
\affil{Queen Mary University of London}
HAMED HADDADI
\affil{Queen Mary University of London}
AKRAM ALOMAINY
\affil{Queen Mary University of London}
}

\begin{abstract}

Health and fitness wearable technology has recently advanced, making it easier for an individual to monitor their behaviours. Previously self generated data interacts with the user to motivate positive behaviour change, but issues arise when relating this to long term mention of wearable devices. Previous studies within this area are discussed. We also consider a new approach where data is used to support instead of motivate, through monitoring and logging to encourage reflection. Based on issues highlighted, we then make recommendations on the direction in which future work could be most beneficial.

\end{abstract}


\begin{bottomstuff}
This work is supported by the Engineering and Physical Sciences Research Council (EPSRC) and the Media and Arts Technology Doctoral Programme. This work was done while Haddadi was at Qatar Computing Research Institute.
Authors' address: School of Electronic Engineering and Computer Science, 
Queen Mary University of London, UK;
\end{bottomstuff}

\maketitle

\section{Introduction}
\label{intro}

Wearable technologies are a relatively new entrant in the health and fitness sector. Examples of health and fitness wearable devices and applications include health monitors, fitness trackers, activity monitors, and analysis aids. According to \citeN{orang2014}, fitness and medical wearables accounted for 60\% of the wearables market in 2013 and it is predicted that the health wearables market will be worth roughly \pounds~3.7 billion by 2019. The ever-growing popularity of smartwatches and fitness bands suggest that the increase of wearable health and fitness devices remains an ongoing trend; the technology associated with these wearable devices is improving at a fast rate. Devices are becoming increasingly smaller and more energy efficient, making them better suited for continuously sensing and giving feedback.

Although the technology is improving and applications are evolving, ensuring long-term user retention is a challenge that still remains. The dropout rate of health and fitness wearable devices currently stands at around 85\%~\cite{1techinasia2014}. The lack of efficient data collection, utilisation and feedback may all contribute to the causes of this issue. Advanced sensors can log  individuals' health data efficiently and can present the users with comprehensive information about their health. However, the meaningfulness of data can have a major effect on a user's behaviour. Poorly presented data or the overload of information can lead to an individual becoming confused and discouraged. This in turn leads to them abandoning their wearable device. 

In this paper, we present an extensive survey of different approaches for data utilisation from wearable and mobile technology with regards to fitness and health behaviour change. We explore examples and studies of wearable sensors, actuators, and applications used to promote health and wellness. We then point out different methods of utilising an individual's data to support positive behaviour change. 

Within this literature review, we focused on wearable sensing technologies. Wearable sensing can be used to provide data on various health aspects. These include body movements, physical activity and behaviour, bio-signals like heart rate, respiration, brain activity, or health influencing environmental factors. A large body of research also focused on the use of mobile phone sensing to pick up those signals, and is therefore included in this survey on the basis that their approaches are transferrable to wearable device applications. A criteria for the presented wearable approaches is the presentation of the collected data to the user, either as raw data, in a processed form, or as persuasive, contextualised feedback. For behaviour change theories and techniques, we mainly focused on the work of the UCL Behaviour change group by \citeN{Michie:2014ua, Michie:2013hw}.

This survey is organised as follows; in Section~\ref{driving-behaviour-change}, we focus on using data to encourage behaviour change through motivation and persuasive techniques, using gaming and social aspects to achieve this. We outline several psychological aspects and theories which surround technology driven behaviour change and present related projects. In Section~\ref{data-representation}, we focus on different approaches of data presentations to influence a person's behaviour. This includes the contextualised and adaptive presentations, as well as cognitive supporting ambient displays. In Section~\ref{support}, we explain a more modern approach where data is used to support and facilitate human health behaviour. We outline studies conducted where data from wearable technology can support intrinsic driven behavioural changes. By analysing studies that have been carried out in these areas, we highlight the challenges posed for the future of wearables in Section~\ref{challenges}. Finally, in Section~\ref{conclusion} we suggest areas and directions for future work that we feel would be beneficial to the field.

\section{Persuasive Wearable Technologies for Behaviour Change}
\label{driving-behaviour-change}

When looking into behavioural change, motivation is a key factor to consider. \citeN{nevid2012psychology} describes the term motivation as the 'factors that activate, direct and sustain goal directed behaviours' \cite[p. 288]{nevid2012psychology}. He further describes \emph{motives} to be the needs or wants that drive behaviours. 

Motivation has the power to cause a person to start more healthy activities but also to continue and repeat these activity routines. If there is a lack of motivation or it is not used in the correct way, this can lead to opposite effects in an individual's behaviour~\cite{arteaga2009combating}. Motivation can be, furthermore, classified into two categories; intrinsic and extrinsic motivation, whereby the latter can be further distinguished in sub-types identified by \citeN{Deci:1985km}. Intrinsic motivation can be seen as motivation created through internal interest and enjoyment, whereby the reward lies within performing a behaviour itself and plays an substantial role in one's well-being. Extrinsic motivation, on the contrary, is generated through external influences and rewards. Extrinsic motivation can be further distinguished into external regulation, introjection, identification, and integration. While external regulation is merely based on external rewards and punishments, introjection already describes a certain degree of internalisation of motivation driven by guilt or a gained feeling of self-esteem; but still inherits an external control. Identification and integrated regulation are more autonomous, self-determined and internalised. \citeN{Ryan:2000jo} describe that there can be a process of internalising originally external motivation when the values are identified as consistent with one's perception of self and the motivation is identified as less controlling. The Cognitive Evaluation Theory of \citeN{Deci:1975gg} specifies factors to facilitate intrinsic motivation through contexts that support the personal feeling of competence. 

\begin{figure}[t]
    \begin{center}
        \includegraphics[width=0.7\textwidth]{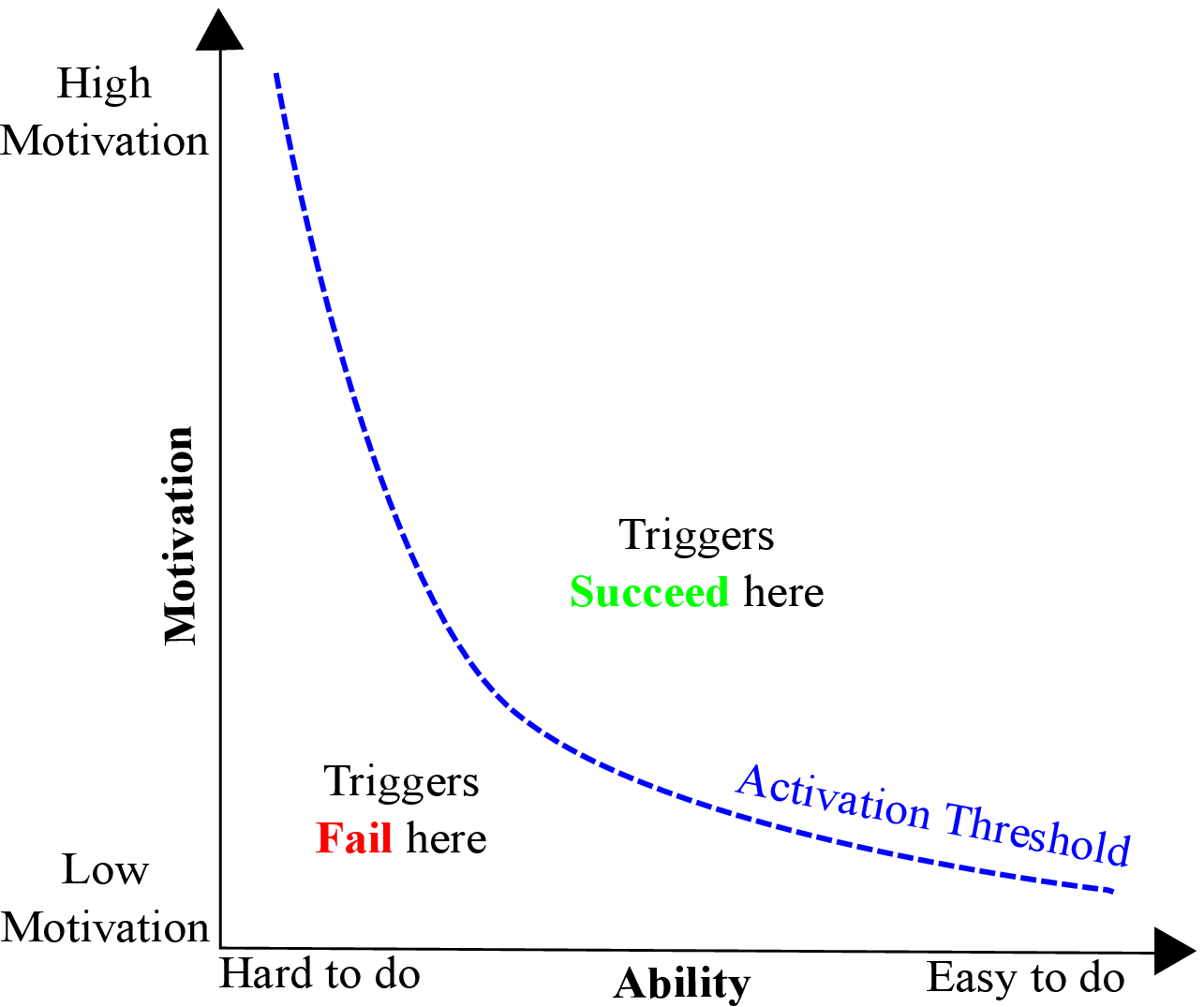}
        \caption{Foggs Behaviour Model (Adapted from~\cite{fogg2009behavior})}
        \label{fig:fogg}
    \end{center}
\end{figure}

Motivation is also an important part of Fogg's \citeN{fogg2009behavior} Behaviour Change Model (Figure \ref{fig:fogg}). It defines three components for successful behaviour: motivation and ability to perform a behaviour in a sufficiently high level and triggers. Different approaches can be made to make a successful behaviour more achievable; this can be accomplished by either making it easier to perform the task by breaking it down in easier actions or by increasing motivation. Gamification aims to highen motivation through use of points, leader boards and rewards created from data. The trigger to carry out behaviour happens whilst the player is in the game itself. Nobody wants to lose and this triggers people to carry out the actions. 

This section outlines two of the biggest data utilisation methods used to create motivation; gamification and social influence.

\subsection{Gamification}
\label{gamification}

Gamification is a common way to motivate behavioural change. It refers to taking game design elements and applying these within other contexts. For example, rewarding an individual with \emph{game points} if they eat healthy food for a day can motivate them and help develop a specific type of behavioural outcome~\cite{deterding2011game}. Gamification is a relatively new concept and the exact origins are unknown. \citeN{pelling2011short} first used the concept of gamification within commercial devices. The main aim of his consultancy \emph{Condura} was to incorporate gaming methodologies into businesses. \citeN{fogg2002persuasive} describes the \emph{Pocket Pikachu}; it is one of the first wearable devices that utilised gamification to become persuasive tool. The simple device includes a pedometer to measure the step count of the wearer. This data is then translated into game points which help the virtual Pikachu to grow. \citeN{Robson:2015ju} state that presenting data and utilising it within a gaming context works well in encouraging behavioural change because it taps into an individual's motivational drivers; particularly intrinsic motivation which is behaviour driven by internal rewards like fun and extrinsic motivation through rewards like badges and game point. Engaging motivation can arise from within the individual because they enjoy the behaviour and experience it as rewarding \citeA{inMotiv}. A comprehensive review by \citeN{Seaborn:2015iz} highlights future directions in gamification research.

\subsubsection {Motivational Affordances}
\label{motivational-affordances}

Wearable applications can utilise motivational affordances and gamification to motivate extrinsically. This includes emotionally positive stimuli like the use of badges, leader boards and challenges. Based on \citeN{hamari2014does}, the points system claims to be the most commonly used method. \citeN{bleecker2007mobzombies} presented the prototype of a mobile based game --\textit{MobZombies}-- incorporating wearable sensors. The sensors provide accelerometer data which is used to move the virtual avatar. Physical movement by the player within the real world are translated into moves within the game. The main aim of the game is to run away from the zombies and collect points through body movement. Rewarding the player with points utilises features that the user is already familiar with and presenting data in this way ensures that the user gains instant gratification and motivation.

An issue that most of the motivational affordances have in common, is the \emph{clouding} of the actual data. Health data is not directly presented to the user and the focus is on the rewards and achievements to stimulate extrinsic motivated behaviour change. The removal of external incentives can lead the the termination of the exercise and termination of health behaviour if the behaviour is not internalised and congruent with personal, self-determined goals~\cite{Deci:1975gg}. 
 This indicates that the behavioural change is strongly connected to the presence of game elements. Further research is needed to determine long-term health behaviour outside of the game world. Furthermore, there is evidence that a special personality type prefers a certain motivational affordance and that applications should take this into account~\cite{Karanam:2014fd}.

\citeN{Payton:2011fp} developed a mobile game to reduce sedentary lifestyles in college students. \emph{World of Workout} motivates the player to increase step count in small amounts throughout the day. The users can define a goal they wish to achieve and the mobile phone application generates suitable \emph{quests} for the player to complete. The user's step count is calculated by using the \emph{iOS shake event} and is then related to the set goal. Rewarding feedback is provided to the user when a goal is reached. The game was found to have a positive effect on participants, with players finding it fun and enjoyable to play. Feedback from other players suggested that they would enjoy a feature to share their data outside of the game. This could include the possibility of posting achievements on Facebook or Twitter. 

Others have also looked at levering extrinsic motivation by enabling the user to earn tangible rewards such as money or psychological rewards like praise. An example of a device that utilises rewards is the \emph{Mymo}. Developed by \emph{Tupelo}\footnote{\url{www.tupelolife.com/}}, the activity tracker allows users to cash in their steps to earn rewards such as mobile talk time and airline miles~\cite{1techinasia2014}. But other studies suggest that the use of extrinsic incentives, like materialistic rewards, may have flaws. 
In general, we can say that extrinsic regulation can negatively influence intrinsic motivation and undermine the feeling of autonomy and competence~\cite{Deci:1985km}; it can therefore have a negative impact on our well-being. \citeN{greene1974effects} stated that when an intrinsically motivated task, such as drawing for children, is rewarded externally, it can be harmful. It leads to people expecting external rewards all the time and can have a detrimental effect on the individual's intrinsic motivations.

\subsubsection{Social Incentives within Games}
\label{social-incentives}

Social aspects are important in gamification. \citeN{ali2006fitster} state that single player games can lead to the user feeling isolated. To avoid this, some gamification strategies include social incentives. There are two main incentives used -- competition and cooperation. Competition can be created by comparing data of two users against each other within an application. In 2010, \citeANP{clawson2010dancing} created a mobile game where players have to dance in time to get points. Two people wear wireless sensors around their ankles that contain accelerometers to measure movements. This data is then translated into game points. Users found this game enjoyable and were satisfied with the experience, and although the participants were just testing the application for two songs, the majority indicated that they would use the game for 20 minutes or more a day. \citeANP{clawson2010dancing} concluded that comparing an individual's data to others within a gaming context can heighten the motivation to dance more and get physical active. 

\citeN{ali2006fitster} discusses \emph{Fitster}, a mobile social fitness application that incorporates large scale competitiveness. It includes an online dashboard which contains the daily step count and activity data of the user's friends. The application accommodates the light-hearted competition that can take place between befriended groups to motivate physical activity. The application is socially oriented and takes this further by allowing users to actively challenge another member to walk a set amount of steps within a given time. The introduction of timing can apply pressure for the user to perform specific behaviours. Although this may not make it as enjoyable, it may increase motivation. 

The second social incentive is cooperation; users can work together to reach a goal and motivate each other within a gaming context. \citeN{ahtinen2010let} created and trialled \emph{Into}; a social mobile wellness application. In the application, the physical activity data from the individuals merges together to achieve a group goal with the aim to encourage physical activity for all group members. The game contains virtual trips between cities in the world made up of step count goals. The player and their group can work together to achieve a goal and gain rewards whilst travelling around the world virtually. The social aspects of the application and the merging of data were found to be beneficial for users.

But which social incentive is the most influential? \citeN{chen2014healthytogether} noticed that many games focus on the competitive element of gaming. They developed a mobile application to find out which social incentive is most influential and observed how players reacted to data presented in three gamification modes. This included competition, cooperation and a hybrid of the two. The application included a messaging service, allowing pairs of users to talk to each other. Users could communicate either to help, support or to taunt one another. Results showed that all modes caused people to increase their daily activity but cooperation was more powerful than competition. Their qualitative results also indicate that users prefer to be paired with a partner with similar abilities.

\subsubsection{Real versus Virtual Worlds}
\label{real-virtual-worlds}

More recently gamification has started merging data from the real world with the virtual world within the game. Mobile games developed by \citeN{macvean2012ifitquest} and \citeN{chuah2012wifitreasurehunt}, created virtual game maps using location and movement data from the phone to bridge the gap between the real and gaming worlds; those games can be considered alternate reality games~\cite{McGonigal:2004vz}. The user must physically move around their physical environment to earn virtual rewards. The audio augmented \emph{Zombies, Run!}\footnote{\url{www.zombiesrungame.com}} game motivates runners by playing zombie noises through the headphones while they are on a run. These noises are supposed to create the immersive feeling of being hunted and motivate the user to run faster. 

\textit{Freegaming} by \citeN{gorgu2012freegaming} is an interactive game using augmented reality through the user's mobile phone camera. It places information and directions over real life footage of buildings and landmarks as the user views them in real time. The aim of \emph{Freegaming} is to motivate outdoor exercising. This is achieved by presenting data about the user's status and environment itself in an immersive, augmented way. Based on this information, the app suggests workout routines to the user. As a result of the study, the author suggests that presenting data within a familiar environment can have an influence on an individual's behaviour. For example, if they know the running route and the rough distance in advance, users may be easier motivated. Getting feedback is an important factor for working successfully towards a goal~\cite{Locke:2002ds}.

\subsection{Social Influence}
\label{social-influence}

 \citeN{Rashotte:2007td} defines social influence as the effect on another humans behaviour, thoughts, and attitudes as a result of interacting with others. 
For example, recommendations by friends likely lead to an desire to try out or buy the same thing. Instances like this sway our own ideas, actions and behaviours on a daily basis. According to \citeN{ledger2014inside}, social factors are important for our health and there is strong evidence to suggest that our behaviours are shaped by the behaviour of our family, friends and even the people we work with.

Social influence can also be a key factor in the adoption of new health behaviours. \citeN{intille2004new} explains how an individual's behaviour can reflect that of their peers. Sticking to a diet can be easier for the individual if they have friends that also engage in healthy behaviours. Having friends that eat bad foods around you can cause unwanted temptations~\cite{Bruening:2012fb}. Various wearable applications offer the functionality to share health data through online social networks. This allows competition or comparisons to happen between all members of the social group. This can lead to members of the group reflecting upon themselves and wanting to change their behaviours~\cite{ananthanarayan2012persuasive}. Effects of social networks can even be observed on the large scale; the effect of social contagion and the propagation of obesity, smoking behaviour etc. can be identified~\cite{Smith:2008ct}. New technologies and online media can even enhance this effect.
There are many different types of social influence which are outlined below. 

\subsubsection{Normative Influence}
\label{normative-influence}

Norms within social groups have influence on our behaviour; \citeN{asch1951effects} explains how normative influence causes an individual to alter behaviour. An individual conforms to a groups social norms in order to be liked and accepted within the group. 
One type of informal conformity is \emph{social proof}~\cite{aronson2007social}; it explains how, in times of uncertainty, an individual observes the reaction of others~\cite{cialdini1987influence}. They will then base their own behaviours upon their observations. \citeN{chang2012food} created a social food journalling application called \emph{Food Fight}. The mobile application allows the user to take pictures of the food they eat and share this with other app users to compare. \emph{Similarity} is one of the factors that increases the effectiveness of social proof~\cite{cialdini1987influence}; we are more likely to be influenced by people we believe to be similar to ourselves. The \emph{Food Fight} app allows to look out for people that have similar goals. The pictures that the users share become part of a timeline called the \emph{food feed}. Within this timeline other users can vote up pictures that they like. As most of the users on the app aim to eat healthy, this means that the most popular images are usually healthy foods. If an individual is unsure of what to eat for lunch they may look at popular up-voted images for ideas. This may influence them to try healthier options if they see it is a popular choice.

Another factor that increases the effectiveness of social proof is \emph{authority}~\cite{cialdini1987influence}. If an individual believes the information to be coming from a reliable and trusted source, they are more likely to conform to it. \citeN{buttussi2008mopet} created \emph{MOPET}, a mobile personal training application. The application utilises real-time fitness data from activity and heart rate sensors; analyse it and provides health and safety advice to the user. This advice comes from a virtual personal trainer which is visualised to the user via a talking 3D embodied agent. The belief that the information comes from a real personal trainer, can have an effect on how influential the information is. The user may be more likely to change their behaviour if they feel the advice is from a knowledgeable source.

\subsubsection{Social Comparison}
\label{social-comparison}

Another type of influence, proposed by \citeN{festinger1954theory}, is the Theory of Social Comparison. It describes, that we evaluate our  own opinions and abilities by comparing them to others around us. This happens to reduce uncertainty and supports an individual learning to define themselves. In the wearable ecosystem, this can mean comparing user data within a group. It can also include representing information in a way to encourage comparison to peers and promote self reflection. \citeN{lin2006fish} utilised social comparison to create a computer game called \emph{Fish'n'Steps}. Within this game, the users wear pedometers which collect their daily step count. A higher step count leads to the growth of the user's animated fish character in the game. Social comparison comes into play when many players place their fish within the same bowl. This encourages the players to look at the growth of the other's fish compared to their own. Presenting the step count data in this manner had a positive influence on the activity levels of the participants.

\citeN{bandura2001social} explains a similar concept of \emph{Vicarious Capability} in his Social Cognitive Theory (SCT); it explains how we do not learn only from our own experiences but from also observing those around us. This also applies to observing others mastering tasks; and the vicarious experience of observing others successfully performing a behaviour can improve the self-belief -- namely our self-efficacy -- in the own ability to master a challenge~\cite{bandura1977self}.
Success of a friend losing weight by using an activity tracker could not just become driver for us ourselves getting the same tracker~\cite{ledger2014inside}, but it could also improve our self-belief to master the same challenge and be successful. \citeN{anderson2007shakra} used a wearable mobile device as a health promotion tool by utilising a groups collected data. \emph{Shakra} calculates the daily exercise levels of users from the measured movement data. These daily exercise levels are then shared amongst the group of friends. The study found that this sharing of data was perceived positively by the participants. It helped them to reflect more upon their own exercise level data and encouraged behavioural change. 
If an individual finds that their friend with a figure they perceive as desirable is more active during the day, this may lead to the friend becoming a role model with an influence on the own behaviour. After reflection, the individual may choose to start walking more in hope to achieve a similar success.

Social comparison has been shown to be a successful driver for health behaviour change, but there can be negative implications on a person's wellbeing and interpersonal relationships. These implications include decreased happiness~\cite{Lyubomirsky:1997hr}, feelings of guilt, and dishonesty to others~\cite{White:2006ix}. 

\subsubsection{Social Facilitation}
\label{social-facilitation}

\citeN{zajonc1965social} describe social facilitation, a type of influence where an individual's performance can be improved by the mere presence of others. This includes working with others within a team or by having an audience. Audience effect within wearable applications may involve sharing fitness data and goals with others. Sharing goals on social networks like Facebook increases the likelihood of an individual changing their behaviour to what they feel is acceptable, because they feel a sense of commitment~\cite{ledger2014inside}. The fear and guilt of letting others down by not achieving goals is a main driver of behaviour modification. \citeN{lim2011pediluma} created \emph{Pediluma} which is a wearable device strapped to the user's foot. It takes the user's step count and maps it to a flashing LED light. The more steps the user takes, the more the device will flash. The ambient manner used to display the data results in the public becoming an audience. The individual may adjust their behaviour to ensure the data presented to the public is promoting a positive self-image. The study found the device to increase the amount of daily steps taken. Public commitment can be important with regards to changing a person's behaviour~\cite{Locke:2002ds}. Contrary, \citeN{fogg2013talk} describes anecdotally how this may not be enough on its own. He owns a pair of scales which tweet out the weight to his Twitter account every time they get used. Although this automated process was supposed to motivate through social facilitation, it was not motivational to him and he did not pay much attention to it. The scales barely attracted his interest and therefore did not encourage him to want to lose any weight. 

\citeN{zajonc1965social} explain another type of social facilitation called co-action which describes the effect on the own performance when other people are carrying out the same task.
\citeN{toscos2006chick} created an app called \emph{Chick clique}. The application aims to motivate teenage girls to exercise more, through use of their fitness data. Data presented in the application levers the power of social relationships to bring about behavioural change. The app includes a leaderboard of each group member's daily step counts. This encourages the girls to talk about health and fitness with each other and allows the application to become a persuasive social actor. Furthermore, this approach can counteract misperceptions of peer thoughts and behaviour, which can be one factor of promoting unhealthy behaviour. Social Norms Theory describes three concepts of discrepancies between actual norms and individually perceived norms: pluralistic ignorance, false consensus and false uniqueness~\cite{Berkowitz:2004vb}. Pluralistic ignorance describes the misperception that the majority of peers behave differently than oneself and can lead to the adaption of less healthy behaviour; for example, a physically active person inaccurately perceives the majority of the social group more sedentary, will likely adapt a less active lifestyle. On the other hand, a sedentary person could falsely perceive the majority of the group as sedentary as well and take this as justification to not become more active -- this is called false consensus. If a person perceives the own behaviour as falsely unique within the group, this can lead to withdraw from the social group. While the majority of studies focus on alcohol consumption and smoking behaviour~\cite{Berkowitz:2004vb}, application which facilitate a raised awareness within social groups can counteract those misperceptions and may promote healthier behaviour within the groups.

Other wearable applications encourage co-action by motivating groups whilst they are physically together. \citeN{mauriello2014social} designed and built a set of wearable, electronic textile displays. These displays support a group of people while they are running. Accumulated running data obtained from sensors is displayed on e-ink screens attached to the back of the runners T-shirts. Their studies concluded that they improved motivation within the group through social facilitation. 
\citeN{karau1993social} also state that group members work harder on tasks if they perceive their contribution as instrumental to the desired team outcome. They will also work harder if they feel their peers are monitoring them.

The use of social awareness can increase the effectiveness of co-action techniques. It can lead to an individual having an active interest in others and how they are doing. \citeN{burns2012activmon} created \textit{Activmon} which is a wrist-worn device. Members of a group all wear the device and it contains a custom-built square LED screen. The device monitors each user's step counts and while LEDs light up on the device to correspond with the progress towards the user's daily goal, the whole group's achievements are also displayed on an ambient screen. This allows the user to see how the group is performing as a whole but also individuals. Displaying other member's data can encourage an individual to become interested in how other members are performing, raise the awareness and sense of team belonging, and support motivation and socially-driven behaviour change.

\subsubsection{Social Impact}
\label{social-impact}

\cite{nowak1990private} describe the theory of social impact which presents three factors that affect the amount of social influence. The first factor is \emph{number of sources} for the influence. As the amount of people providing data increases, so does the influence exerted on each individual. This is evident in traditional support groups. 
The second factor is \emph{strength}, which refers to the perceived importance of the feedback source. The more trusted the source providing feedback are to someone, the more likely they are to influence them. As mentioned earlier by \citeN{lin2006fish}, \emph{Fish'n'steps} placed many individual's fish in the same bowl. Family members' fish were more influential than strangers' when placed in the same bowl.

The last factor is \emph{immediacy}, the closeness of the group both in time and space. Wearables open up opportunities for people to share data with others and be within the same shared digital space. Online social networks provide great opportunities to create these digital spaces. \citeN{lu2014reducing} developed an app called \emph{UOIFit}. The mobile application aims to increase levels of activity amongst adolescents. The app collects fitness data of each user and shares this data to everyone through a \emph{fit feed} tab. The app also allows users to exercise with each other remotely. This is an example of creating digital spaces for collaboration through data. Studies conducted into the app found social aspects to have a positive impact on an individual's behaviour. It increased the users' activity level and lowered their Body Mass Index (BMI).

\subsubsection{Wearable Computers as Social Actors}
\label{social-actors}

\citet{fogg2002persuasive} talks about wearable technology becoming a social actor itself and outlines the possibility of communicating with a wearable device to create a social experience.
Wearables themselves may be able to lever social influence. This social influence may persuade individuals to change their behaviours. Fogg highlights cues that can lead to a wearable becoming a persuasive social actor, the first of which is physical cues. 
The more attractive the interface of a device is to the user, the more of an impact it will make on the user (e.g.~\cite{Sonderegger:2014hb, Chang2014168}). The way that data represented aesthetically makes a great difference with regards to influence. 

The second cue is psychological cues. This involves making the user believe that the application possesses emotions and feelings like a human. A subgroup of the Affective Computing community focuses on the technology imitation of human emotions~\cite{Picard:1997:AC:265013}. As \citeN{lin2006fish} showed, \emph{Fish'n'steps} represents the collected activity data in the form of a fish avatar growing. This representation works well in developing a persuasive, emotional relationship between the user and their data. But as found in the study it can also have a negative effect on the user's behaviour. The user would feel responsible for the avatar when it was not growing or looking sad. This feeling of guilt would lead some people to avoid the application as they did not want to see an unhappy avatar. 

Language is another cue mentioned by \citeN{fogg2002persuasive}, this can involve the application asking questions and offering a dialog. Wearable applications can offer praise, which is common among health and fitness applications. The wearable collected data can be used to determine when praise should be given. Using data to create a conversation with the user is important in influencing long-term behaviour change. \citeN{Arteaga:2010} created an application which uses a talking head to communicate motivational phrases and advice to the user at appropriate times. The use of a talking head made the application more anthropomorphic. This was perceived positively in studies and found to make a difference to the users. \citeN{bickmore2008negotiating} conducted studies looking at virtual agents and how they were able to influence health behaviour change in more depth. The virtual agent talked to the user, but they also utilised non-verbal behaviour and facial expressions to communicate messages. These messages were based on data collected from the user. The agent was part of an application designed for use in the office, and the application's virtual agent suggested short breaks to the user at timed intervals. Their studies concluded that the more social cues were presented by the agent, the more breaks the users were taking. Visualising data to a user in an empathetic way may encourage long-term compliance with new behaviours.

\section{Wearable Data Representation}
\label{data-representation}

In current wearable applications, visual representations and methods of feedback are the most common way to communicate data to the user \cite{ludden2013designing}. The way in which the raw data is processed and manipulated before it is presented to the user, plays a big role in how influential it can be. One of the issues raised by researchers in the health and personal data space is provision of meaningful data to the users and enabling Human Data Interaction~\cite{haddadi2013human}. 

Current step counters, pedometers and their accompanying applications struggle to provide long-term behaviour change. The users are still aware of the data provides of these devices, but it is not perceived as so meaningful and influential after six months~\cite{ananthanarayan2012persuasive}. Data representations need to emphasise the importance of being healthier in more intuitive and meaningful ways.

\subsection{Adaptive and Contextual Data}
\label{adaptive-contextual}

Adaptive and personal data representations can be a powerful tool to deliver meaningful information from wearable devices. An individual's health and fitness is a personal issue to them, so the way of presenting the data should be just as personal. By making the representations user adaptive, a personal experience for each user can be offered. 

A context aware system is able to tell what the user is doing from utilising and analysing sensor data. It can then use this to make motivational suggestions at specific, influential times in the day. \citeN{intille2004new} conducted a study that looked into utilising just-in-time messaging of health information. The presentation of contextualised information at important decision times within the user's day proved to be effective and previous studies show that just-in-time persuasive interfaces can influence behavioural change. \citeN{mitesh2015wearable} suggest supporting new behaviours may be best facilitated by regular, appropriately framed feedback. This feedback should be presented at the times where the user is most likely to notice it. \cite{carroll2013food} looked at a sensor-equipped bra that senses emotions. They plan to utilise the collected data to determine when the individual is most vulnerable to emotional eating and then present just-in-time interventions. \citeN{intille2004new} had four rules which they believed helped the data representation to be effective. The first rule was to keep the data representation as simple as possible. This ensures the user understands the data that's displayed to them as clearly as possible. Other rules include displaying data at appropriate times and in the appropriate place. This is to make it as easy as possible for the user to refer to the data within their day-to-day life. If the data representation is irritating to the user this can lead them to ignoring the device altogether. 

\citeN{gockley2006aviva} created a wearable device that contextualises sensor data. \emph{Aviva} tracks the users' and their close friends' eating and exercise patterns. A wrist-worn, watch-type device shows the feedback to them. It aims to display qualitative and holistic data to the user, not just simply numbers from sensors. Experts suggested that displaying lots of raw unexplained data can lead to the user becoming discouraged \cite{gockley2006aviva}. \emph{Aviva} displays personalised, contextualised suggestions. For example, the user could be notified to 'eat a bag of nuts' as opposed to just telling them they 'need more protein'. This contextualising of data can be more persuasive when getting an individual to change their behaviour. Another way to contextualise data demonstrated by \citeN{macvean2012ifitquest} and \citeN{chuah2012wifitreasurehunt} in 2012. They suggest to give recommendations specific to the user's location. Making an application location aware can lead to it being more entertaining for the user as they feel they can relate to the data more easily.

Previous studies have shown personalisation to be an important factor for device usage and adaptations. Studies outlined by \citeN{ananthanarayan2012persuasive} have shown that users want to create their own system to access their data. Users do not like having to use a predefined default form that everyone else uses as it does not feel personal. \citeN{ananthanarayan2014towards} developed another wearable to look into the effectiveness self-crafted devices. The device the users could create held a set of sensors that tracked an individual's UV exposure. They found that people received the device well with people attaching it to their bags or even wearing it as an headband. Another device developed by \citeN{ananthanarayan2010health} looks into raising health awareness in children. They wanted to motivate them to think about their behaviour. It had a personalised build-it-yourself approach to the fabrication of the devices. All the components the device contained were \emph{plug and play} in style. The children could attach components to a leather bracelet, including sensors and displays for feedback. The LED display changed colour dependant on the amount of exercise the child had done that day. Allowing individuality for data representation and form factor may make the device more meaningful to the user. This is because they feel they have created it so it may be more influential than a standard generic form. 

\subsection{Triggering Habits}
\label{subs:habits}

Presenting meaningful data to the user is encouraging in promoting behaviour change, but presenting it in a way to compliment behavioural change processes is the most effective method to use. 
\citeN{fogg2013talk} suggests behavioural change is systematic. He also suggests that motivation is not enough to sustain long-term behaviour change. Fogg created \emph{Tiny Habits}, tiny life changes that become automatic. For example he decided to do two push ups every time after he went to the toilet. He increased this slowly everyday until he was able to do 20 pushups a day. Wearable applications and the presentation of data should not place importance on motivating the individual. They should instead accommodate letting the natural processes emerge. He quotes that an application can \emph{``plant a seed in the right spot and it will grow without coaxing"}.

Habits play a substantial role in forming behaviour. As a "behaviour that has been repeated until it has become more or less automatic, enacted without purposeful thinking, largely without any sense of awareness" \cite[p. 1]{Nilsen:2012ju}, habits are hard to consciously establish and to change and require self-consciousness, willpower and self-regulation. According to \citeN{Strack:2004bz}, motives and behaviours can be addressed to two systems; the Reflective Impulsive Model describes these two two systems as the reflective system and the impulsive system. While the reflective system is characterised by cognitively intensive decision processes based on knowledge and active willingness to perform the behaviour, the impulsive system influences actions concerning basic human needs and habits; and it performs with less cognitive load for the triggering of these actions. Technologies which support the information processing with less cognitive load could therefore lead to an easier suppression of habitual behaviour in the right moments. \citeN{Strack:2004bz} also highlighted is the role of the physiological arousal on the reflective system. While a high level of arousal promotes the execution of habits and stereotypic behaviour, low levels of arousal weaken self-control. A mid level of arousal appears to be the optimal level for the reflective system and for addressing conscious behaviour change. Context aware sensing to detect stress and arousal could lead to context-aware systems which could give recommendations in the right moment. \emph{Food and Mood} by \citeN{carroll2013food} considers first approaches in this direction to intervene emotional eating habits. Other projects looked into the identification of boredom using mobile phone sensing \cite{Pielot:2015ee}. These approaches could build the basis for future contextualised and emotion aware interventions.

The establishment of habits plays a substantial role in establishing long-term behaviour success. When we look at the Transtheoretical Model from \citeN{Prochaska:1982jn}, the transition from the action stage to the maintenance stage includes a longer-term adaption of the new behaviour, an increased self-confidence and less likelihood to fall back into old behaviours. This is also a prerequisite for habits

\citeN{rajanna2014step} created an application called \emph{Step up life}. The application uses the suggestion of small contextually suitable activities at regular intervals. \emph{Step up life} promotes brief bursts of physical exercise after periods of inactivity. It does this by sending data in the form of on screen nudge reminders. Suggesting small behaviour changes to the user can lead to the incorporation of these changes into their daily lifestyles. The changes seem easy to do by the user so they are more likely to lead to long-term behaviour change than setting up unrealistic goals.

\subsection{Theory of Planned Behaviour}
\label{planned-behaviour}

\begin{figure}
    \begin{center}
        \small
        \includegraphics[width=0.7\textwidth]{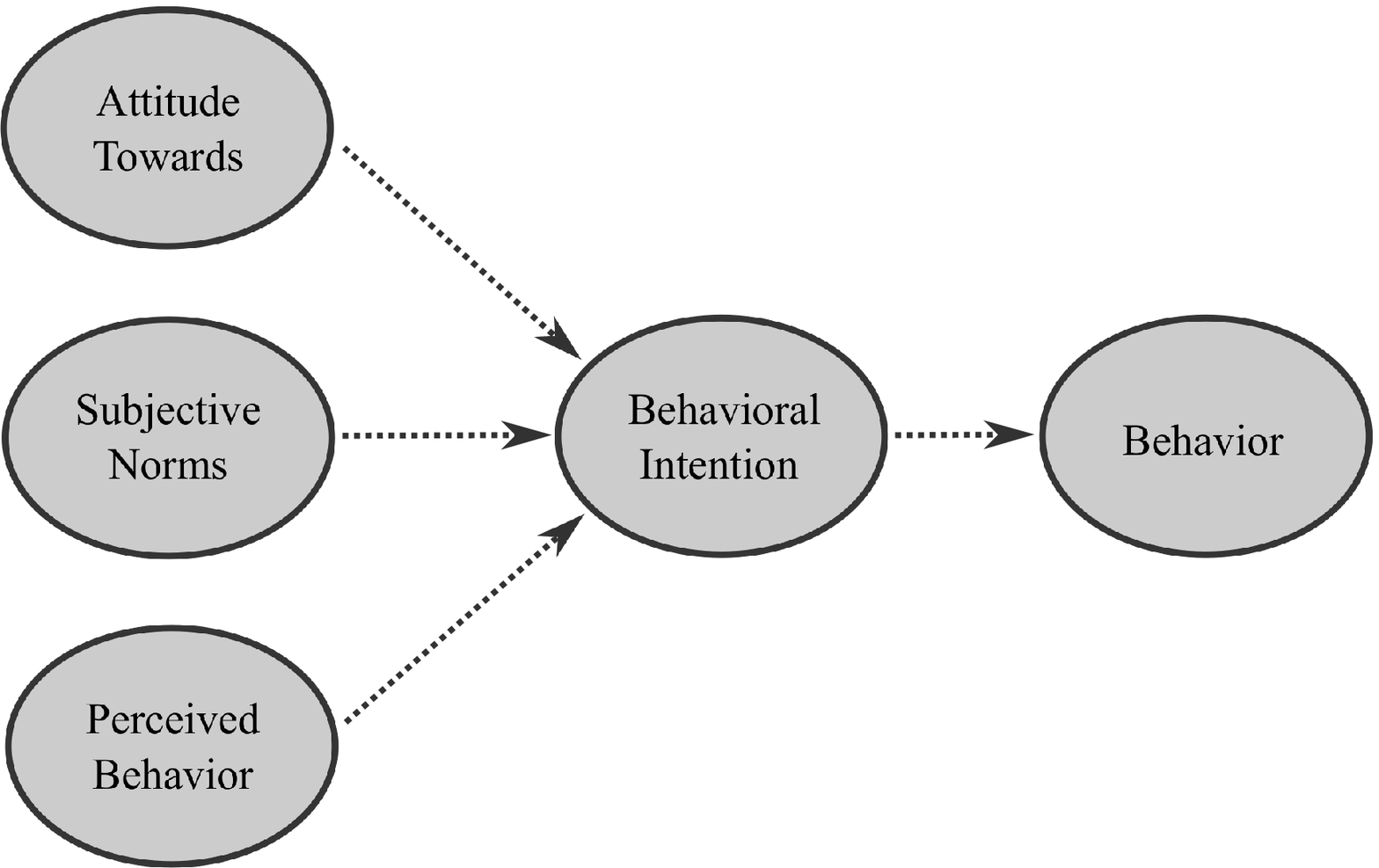}
        \caption{Theory of Planned Behaviour}
        \label{fig:planned}
    \end{center}
\end{figure} 

When thinking about how best to personalise data representation to make it as persuasive as possible, the Theory of Planned Behaviour (TPB) could be considered (Figure \ref{fig:planned}). Created by \citeN{ajzen1991theory}, TPB states that intentions are the best predictor of how an individual is going to behave in certain situations. For example, if we plan to do something we are more likely to go through with it. Three factors produce an individual's intentions to perform a specific behaviour. These are their attitude, subjective norms and perceived behavioural control. Behavioural attitude refers to how the individual feels about the behaviour. This includes affective attitudes which describe whether they feel they would enjoy doing it. Also, instrumental attitudes refer to whether they feel a behaviour would benefit them. Contextualising health data to make the user aware of the benefits may have an impact on the user's behaviours. Subjective norms deal with the support that we get from our friends, family and even the doctor. Injective norms involve others encouraging specific behaviours. An example of this is a friend making you go to the gym. Descriptive norms involve others actually engaging in a specific behaviour. This would involve your friend actually going to the gym with you. Wearable ecosystems have adapted the social aspect of sharing data to address these social norms. The final factor is the extent to which the individual believes that they can carry out the behaviour. This is influenced by how the data is presented to them. If a task sounds easy, an individual is more likely to engage in it.

\citeN{Arteaga:2010} developed a mobile phone application that considers TPB within its design. The app aimed to change an adolescents behaviour and get them to exercise more. They wanted to achieve long-term behavioural change as opposed to the short-term changes that other fitness games were producing. To design data representations that are engaging for teenagers, the system incorporated TPB principles. To stop the user getting bored of the game they decided it needed to take their personality into consideration. Everyone's preferences are different so they created a game that assesses the person's attitudes. Based on this assessment, it would suggest games that would be the most motivational and beneficial to them. Adjusting data representations around an individual's attitudes and personal traits can lead to stronger intentions for behavioural change. 

\subsection{Ambient Displays}
\label{ambient-displays}

We mentioned before that the way of presenting the data from wearable sensors to the users plays an important role in supporting long-term health promotion. The displaying of numbers and figures may not be enough to encourage behavioural change.
\citeN{ananthanarayan2012persuasive} state that wearable technology has the capacity to track difficult metrics such as heart rate. The right presentation of this data to the user can have an impact on the influence on behaviour  change. The subtle and ambient presentation of the data can be a key to motivate and subconsciously trigger a behaviour change. 

When presented with a lot of complex information an individual can lack the cognitive capacity to process it. \citeN{ham2010ambient} show that ambient displays can be more persuasive because they do not require the user's conscious attention and use little cognitive resources. Studies conducted have found that the use of simple displays for information have more effect on an individual's behaviour than displaying numerical values. Modern health wearable devices have started to harness the power of using ambient ways to display data. \citeN{consolvo2008activity} developed an early example of such an application. \textit{Ubifit} is a mobile application that includes a glancable display. This display is a non-literal representation of the physical activity that the user has done. It also represents the goals that they have achieved. The display contains a metaphor of a garden and the user gains more flowers by exercising more. They gain butterflies in their garden for achieving goals. The application collects data from sensors on the user's body. It then analyses the data to change the aesthetics of the garden display throughout the week. From studies they discovered that participants did find the display motivating. They agreed that the metaphorical representation of their data helped them to focus on their goals. 
\citeN{Lin:2012jz} presented another similar application called \textit{BeWell+}. The metaphor of fish within a fish bowl is used. The more fish it shows, the more physical and socially active the users have been. It also gives feedback on sleep quality by changing the light in the underwater world. This application gives unobtrusive feedback whenever the user glances at the screen of the phone. This subconsciously promotes healthier behaviour and wellbeing. \citeN{fortmann2014waterjewel} created \textit{Waterjewel}, a wrist worn device which aims to influence the user to drink more water throughout the day. It has a light up display that indicates to the user how much of their daily goal they have already achieved. The device also flashes every 2 hours as a nudge to tell the user to drink more water. Studies conducted with the device found it was successful in promoting healthy drinking behaviour. Users did drink more water when wearing the bracelet then when not. 

\subsection{Cognitive Dissonance}
\label{cognitive-dissonance}

\begin{figure}
    \begin{center}
        \small
        \includegraphics[width=0.9\textwidth]{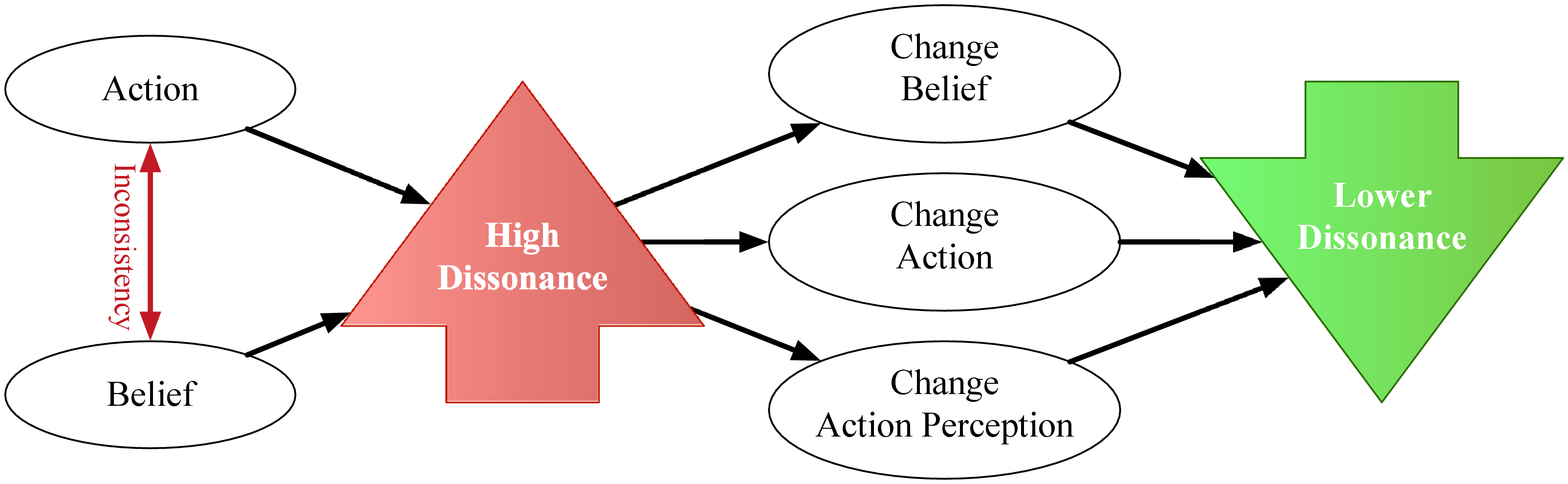}
        \caption{Cognitive Dissonance Theory}
        \label{fig:dissonance}
    \end{center}
\end{figure} 

When thinking about the reasons that ambient ways of data representation are effective, cognitive dissonance theory (CDT) offers some insight (Figure \ref{fig:planned}). Created by \citeN{festinger1962theory}, cognitive dissonance theory refers to situations that cause a conflict for an individual's attitudes, beliefs or behaviours. As humans we have a inner drive to keep all three of these in harmony. Contradiction can lead to discomfort which causes the individual to change their behaviour to restore the balance. Using displays and feedback techniques can also present data in a way to get the user to think of the long-term effects of their current short-term actions and cause the contradiction effect. 

The \textit{Fatbelt} created by \citeN{pels2014fatbelt} is an example of CDT in action. The device looks into utilising isomorphic feedback to get the user to think of the consequences of their behaviours. The user wears the device around their waist. It uses physical feedback by inflating around the user's stomach when they consume too many calories. This simulates the long-term weight gain associated with overeating. In tests the device contributed to a significant decrease in calorie consumption from the user. The use of data in this way leads to the user feeling that the device is an extension of their own body. This gives the wearable more emotional power over the user and their behaviour. 
\citeN{zhang2013see} created a similar device which uses augmented reality glasses to represent potential UV damage on the user's skin. It discouraged them from staying out in the sun too long as was also found to encourage healthy behaviour.

Both techniques made the users more aware of the future consequences of current unhealthy behaviour. This makes them feel uneasy about continuing the behaviour. This raised awareness on problematic behaviour and the focus on consequences can be an initial trigger to start a behaviour change. \citeANP{Prochaska:1982jn}'s Transtheoretical Model \citeyear{Prochaska:1982jn} describes stages of behaviour change and that different techniques can support people to change in different changes. Technologies which focus on consciousness raising can support a transition from the pre-contemplation stage, where people are not considering a behaviour change in the foreseeable future, to the contemplation stage, where people are aware of the problem and are seriously considering to change the action in the foreseeable future (the next six month) \cite{Prochaska:1982jn}.

\section{Support of Behaviour Change}
\label{support}

Methods of using wearable device applications to drive behaviour change have proven to work well in previous research. But the feasibility of such methods in ensuring long-term retention is an area that needs further research. Recent studies focus on using the data that wearable technology provides to support individuals with behaviour change. 
\citeN{fogg2013talk} showed concerns about technology focusing on \emph{motivation} for behaviour change for the western culture. He believes that systems that \emph{support} behaviour change would be much more successful in the long term. Similarly, \citeN{WatchPromotingMot} states in his TED talk: 'Don't ask how we can motivate people. That's the wrong question. Ask how we can provide the conditions within which people can motivate themselves.' Wearable products are starting to support instead of drive behaviour and are utilising data with the aim of influencing our inner abilities to change behaviour. 

\subsection{External and Self-Monitoring}
\label{monitoring}

This section focuses on the use of wearable technology to monitor health and wellbeing. These aspects can be monitored by the users themselves or by an health professional. We briefly outline sensing technologies used to provide health related data. This information about our behaviour can be a prerequisite for enabling behaviour change. Fogg's behaviour change model describes 'ability' as factor of behaviour~\cite{fogg2009behavior}. Monitoring devices can provide information and ease our ability reason about behaviour. We then focus on the use of data for self-reflection, which can help improve self-understanding. This increased self-understanding can lead to an individual making changes and informed decisions in their everyday behaviours. Similarly, COM-B by \citeN{Michie:2011ce} model identifies the \emph{capability} as an important component to enable behaviour. Capability can be seen as the psychological and physical capacity of a user to engage in a behaviour; and it is partially characterised by being able to comprehend and reason about the target behaviour. This comprises knowledge about the behaviour as well as the ability to compare the behavioural performance with the target behaviour. Behaviour monitoring is an important part in this process to enable users to make informed decisions. 

\subsubsection{Health Monitoring}
\label{health-monitoring}

The increasing accuracy and portability of health monitoring sensors is promoting less obtrusive data collection and enables long-term health monitoring~\cite{Pantelopoulos:be}. There are many examples where wearable technology is used successfully in monitoring an individual's recovery from illnesses and rehabilitation~\cite{Patel:2012jx}. While this health data can be used by the user to adjust heath behaviour, the data provided by wearable devices allow this monitoring to happen remotely, too. An example of this could include a medical professional being able to monitor patients without them having to be in the hospital. This is advantageous because the patient can benefit from healing at home. Being at home is more comfortable for many and this can lead to improved healing compared to a recovery in hospitals. Remote monitoring also cuts costs for the healthcare system due to shorter hospital stays.

\newcolumntype{S}{>{\raggedleft\arraybackslash\hsize=.9\hsize}X}
\renewcommand{\arraystretch}{2}
\begin{table}[h]
    \tbl{Common wearable sensors and example applications\label{tab:sensors}}{
    \begin{tabular}{@{}>{\raggedleft\let\newline\\\arraybackslash\hspace{0pt}}p{0.13\textwidth}@{\hspace{0.02\textwidth}}p{0.43\textwidth}@{\hspace{0.02\textwidth}}p{0.4\textwidth}@{}}
        \bf\centering Sensor &  \bf\centering Measurement &  \bf\centering Examples \tabularnewline
        \cline{0-2}
        Accelerometer &
        Is usually used to determine movements and activity by measuring the acceleration.  &	
        \begin{itemize}
            \item Activity monitor: \cite{clawson2010dancing, Bulling:2014jm} 
            \item Movement execution in rehabilitation: \cite{How:2013et, Nerino:2013fd} 
            \item Habit tracking (smoking, food intake): \cite{LopezMeyer:2013go, Amft:2005cq} 
            \item Sports performance: \cite{Spelmezan:2009jy}
        \end{itemize} \tabularnewline
        \cline{0-2}
        Stretch sensors (textile) & 
        Stretch sensors are flexible sensors that change conductivity when stretched or bend. & 
        \vspace{-4mm}
        \begin{itemize}
            \item Measure angle of joints in rehabilitation: \cite{Shyr:2014cm}
            \item Movement of the chest to determine respiration rate: \cite{Qureshi:2011tn}
        \end{itemize} \tabularnewline
        \cline{0-2}
        Piezoelectric sensors (textile) &	
        Piezoelectric sensors measure force/pressure applied to them &	
        \vspace{-4mm}
        \begin{itemize}
            \item Tracking of hits Taekwando \cite{Chi:2004gp} 
        \end{itemize} \tabularnewline
        \cline{0-2}
        Heart Rate sensor (ECG or PPG)  &
        Heart Rate sensors can be used to measure the activity of the heart, which gives indication on health, energy expenditure or arousal levels. It can be measured with Electrocardiograms (ECG) or Photoplethysmogram (PPG) & 
        \vspace{-4mm}
        \begin{itemize}
            \item For more accurate calculation of Energy Expenditure: \cite{Altini:2013wj} 
            \item Fetal monitoring with special belt: \cite{Fanelli:2012tg}
        \end{itemize}  \tabularnewline
        \cline{0-2}
        UV sensors	& 
        Sense the amount of UV light & 
        \vspace{-4mm}
        \begin{itemize}
            \item For warnings when there is too much sun light exposure: \cite{zhang2013see}, \citet{ananthanarayan2014towards}
        \end{itemize}  \tabularnewline
        \cline{0-2}
        GPS &	
        GPS is used for localisation &
        \vspace{-4mm}
        \begin{itemize}
            \item Used to contextualise other data: \cite{macvean2012ifitquest} and \cite{chuah2012wifitreasurehunt}
        \end{itemize}
    \end{tabular}}
\end{table} 

Wearable devices bring a wide variety of sensing to detect the amount, type and execution of movements. An overview of sensors and their application can be found in Table~\ref{tab:sensors}. Accelerometers are devices which determine acceleration data and can therefore detect movements. These sensors are widely used for activity tracking~\cite{Garcia-Ceja2014}. They can be present in either in mobile phones or body-worn devices. Accelerometers form the basis for data collection in many commercially available activity trackers. Utilising Accelerometer data can identify activities such as walking, running, eating and drinking movements~\cite{Amft:2005cq, Bulling:2014jm}. They can even detect human behaviours such as smoking~\cite{LopezMeyer:2013go}. A clinician or the individual themselves can analyse the automatically recorded data. Analysis of behaviours and habits can help support a healthy behaviour change by forming a basis for services such as counselling or other intervention methods. 

This also brings new possibilities to the physiotherapies after injuries or surgeries. \citeN{How:2013et} created a mobile application called \emph{MyWalk}. The application supports patients who have suffered from a stroke in the past. Step patterns of the wearer provided from the phones accelerometer detect gait asymmetry. If asymmetry exists, it lets them know that they need more training to establish a symmetry in their step pattern again. The mobile app offers different trainings modes and a overall score after each session. The user is able to view their score history to review their improvements. They can also share this score history with their therapist. The data collected from their training at home may help to enhance their physiotherapy sessions and enables patients to monitor a part of their rehabilitation from home. \citeN{Nerino:2013fd} focused on the rehabilitation after knee surgery. They used accelerometers to collect data at different positions around the leg. They then used this data to monitor motor functions of the exercising patient. They created an application that included a coaching function, which would suggest exercises. There is also a video conferencing functionality for situations when the therapists is needed. \citeN{Patel:2012jx} presents a detailed review of wearable sensors that are currently used for rehabilitation and he especially identifies the trend of using ambient sensing for holistic home health monitoring and the need for a telepresence integrated in home monitoring systems.

Textile sensors and fabrics are other enablers for wearable technology in the healthcare sector. The combination of conventional, non-conductive fabrics with conductive materials have led to new sensor technologies~\cite{Marculescu:2003fk}. These new technologies allow easy integration into textile products and garments. Stretch sensors are an example of these new technology. These sensors are able to collect data that can be used to monitor movements of joints in the body~\cite{Shyr:2014cm}. \citeN{Qureshi:2011tn} used knitted stretch sensors to monitor breathing and \citeN{Rai:2013em} used textile sensors to monitor neurological and cardiovascular biosignals. Textile sensors bring the possibilities of flexibility and unobtrusive integration into clothing. Where delicate and soft sensors are required, flexibility can be an advantage. An example of where this is important is textiles for newborn infants. \citeN{Chen:2010cv} developed neonatal babywear that measures the temperature of babies using soft textile sensors. The sensors were designed to be aesthetically pleasing but also as comfortable as possible for newborns to wear. An extensive review of wearable, smart textiles has been presented by \citeN{Stoppa:2014km}.

Another use case is the monitoring of health parameters to give the patient peace of mind and contact a medical professional in emergencies. Wearable sensors can be used to monitor pregnant women. \citeN{Fanelli:2012tg} monitored the fetal heart rate with a stomach belt. The designed the belt to be easy to put on to ensure it is easily useable at home. This reduces hospital visits during the pregnancy and make the pregnant woman feel calmer.  

There are several technologies for detecting seizures with wearable technology which could provide data to inform clinicians or family of a seizure. \citeN{Patel:2009kn} used accelerometer data and EEG brain signals to detect seizures with a 95\% accuracy. The \emph{Human+} platform created by \citeN{Altini:he} uses various sensors like EEG, heart activity via ECG and skeletal muscle activity via EMG to obtain data and detect seizures. While these approaches are not very usable in everyday applications due to the use of EEG and ECG electrodes, the MIT Affective Computing group developed a seizure detecting wristband which uses Electrodermal Activity~\cite{Poh:2012jj}. This research formed the foundation for the commercially available Embrace watch\footnote{\url{https://www.empatica.com/product-embrace}}. The watch has an accompanying app, which alerts parents or caretakers in the event of a seizure of the child or patient. Additionally it can be used to monitor stress and sleep levels. Wearable technology can be used to monitor behaviour and use the data to gain insight into health states of a person. \citeN{Madan:2010fg} looks at the usage of mobile phone data to detect the health status of an individual. The data is analysed to detect health conditions such as colds or depression. This application can form the basis for informing the user's doctor of their condition. More broadly available mobile phone sensing can be used for epidemiological studies amongst large populations.

In healthcare it is often necessary to avoid unhealthy situations. Data provided by wearable technology can help identify unhealthy situations and environmental influences. One example is the UV sensing glasses from \citeN{zhang2013see}. Too much sun exposure is widely known to be connected to skin cancer. These glasses keep track of the sun exposure and warn the user when they are at risk. \citeN{Fabrizi:2014jp} presented a concept for a wearable textile flower. The flower is a visual representation of air quality data and can raise the awareness about unhealthy polluted air. 

\subsubsection{Quantified Self}
\label{qs}

The Quantified Self (QS) is a new movement supported by sensing data obtained from wearable devices. QS is part of the Personal Informatics, focusing on tools to support the personal growth and improvement and an individual. Within the QS community, this is mostly achieved through the use of technology in data collection and analysis with the focus on collecting data about ourselves with the purpose to reflect. Reflection can increase self-understanding about areas that need improvement in the future~\cite{Swan:2013uh}.

\citeN{Choe:2014bo} state that health improvements are one of the most reported reasons for self-quantification and especially \emph{activity} is a commonly tracked feature. Commercially available fitness trackers like \emph{Fitbit} or \emph{Jawbone}\footnote{\url{www.fitbit.com/} and \url{www.jawbone.com/}} allow daily assessment of steps and activity. Fitness watches, like the \emph{Atlas}\footnote{\url{www.atlaswearables.com/}} promise the automation of workout logging, which is usually a manual task. It achieves this by using data to identify workout activities and repetitions. The consumer market for self-tracking wearable technology is ever growing. Tools like the open-source \emph{Fluxtream}\footnote{\url{www.fluxtream.org/}} support self-tracking by providing a platform for data aggregation and visualisation from multiple sources. It also supports the identification of correlations within the data. 

Quantified Self can be a powerful tool to gain insights and support behaviour changes within an individual by providing information on the own performance; this allows users to identify . But it can requires skills to understand data fully to ensure long-term engagement. \citeN{Choe:2014bo} identifies two reasons why self-tracking often fails. One is that too many things are tracked and the effort is relatively high to track these. Automated data capturing and simple tracking mechanisms can ease the burden of tracking. The second reason is lack of knowledge about triggers and the context of the data. This confusion around the interpretation of data makes behaviour change difficult for the user. Self-tracking can be a powerful tool for reflecting and making us more aware of our own daily habits, patterns and performance. But at the current time, it still requires a lot of effort, engagement and knowledge. Creating tools to ease these hurdles could assist us by identifying our own behavioural and habitual patterns. By presenting this data back effectively it could provide active support for self-improvement.

\subsubsection{Sports Performance Monitoring}
\label{sports-performance}

When it comes to sports, performance monitoring is essential in improving performance or preventing injuries. Using data monitoring can be useful for individual or even group performance. \citeN{Strohrmann:2011gq} looked at the use of shoe sensors to access kinematic parameters of runners. This data can give an insight into the runner's performance and technique. This can help medical professionals and the runners themselves to analyse how effective the training was. The \emph{Sensoria} fitness tracker works in a similar way\footnote{\url{http://www.sensoriafitness.com/Technology}}. 
The \emph{Sensoria} tracker consists of a sock with textile pressure sensors and an attachable main unit for the data transfer to a mobile app. It can provide data about the performance during a run as well as feedback about the right running technique. 

\citeN{Spelmezan:2009jy} looked at the use of force, bend and accelerometer sensors to track the movements of snowboarding beginners. The data collected is then shared with their trainers. The data helps the trainer to give more accurate feedback on the movements and technique of the snowboarder. They suggest a system like this could help in the teaching process. \citeN{Chi:2004gp} looked at the tracking of movements of Taekwondo players. They used piezoelectric sensors to detect forces applied by hits of the competitor. The system then counts the hits and calculates a score based on this information. This calculated data is then used as feedback for the athletes, trainers and the jury in competitions. 
Not just the performance of an individual is important. In team sports, the performance and communication of the whole team matters. Technological advances within networks and algorithms allow real-time assessment and remote monitoring of bio-signals within a group of athletes. \citeN{Garcia:cn} gives an example for using bio-signal sensing to monitor a group of soccer players.

\subsection{Encouraging Reflection}
\label{reflection}

Reflection is uniquely human ability and belongs to the basic capabilities defined in the Social Cognitive Theory by \citeN{bandura2001social}; next to our capability to abstract, to learn from others, think about the future, and validate our own behaviour against standards. While reflection is a cognitively heavy process, it involves introspection and a willingness to learn; and allows us to make decisions based on knowledge~\cite{Strack:2004bz}. Reflection plays a substantial role in all the above presented sections on external and self-monitoring. 
Getting users to reflect upon their own behaviour can help them to stay on track by supporting their learning and introspection; and it builds the basis for to self-control. It helps us identify discrepancies of our own behaviour to the desired outcome and supports the regulation towards goals. 

\citeN{Fleck:2010cs} distinguished different levels of reflection, these include revisiting, revisiting with explanation, exploring relationship, fundamental change, and wider implications. The Quantified Self movement and the exploration of relationships 

\citeN{Fleck:2010cs} present an overview on technology support for reflection. One support type focuses on helping us revisit information, moments and thoughts from the past. While they consider 'revisiting' not as a reflection process, it is a facilitator. Technology which records aspects of our lives and let us revisit these aspects later, can become such a facilitator. Lifelogs can be seen as such technology; the aim of a lifelog is to record various aspects of a person's lifestyle. \citeN{gemmell2006mylifebits} created one of the first technologies to support lifelogging. \emph{MyLifeBits} creates a complete historical log of documents, websites, and other objects a person has encountered whilst using their computer. To build upon the effectiveness of presenting data in a lifelog, \citeN{epstein2014taming} utilised the \emph{Moves} mobile phone app\footnote{\url{www.moves-app.com}} which records activity and location data on the phone. They developed \emph{cuts} which focus on a subset of the data with a shared feature, e.g. days with the most physical activity or time to commute by the type of weather at the day. New visualisation offered new insights in the data for the user and were perceived positively.

Technologies can also support gaining explanations for experiences or on data. In section \ref{health-monitoring} we highlighted health monitoring systems which allow the sharing of data with health professionals. This dialog can support the process of making sense of the own health data. \citeN{Kocielnik:2013dt} focused on the long-term stress monitoring at the workplace. They used wristbands to monitor stress levels and combined it with data collected from online calendars to generate an aggregated view for self-reflection. This allowed workers to review the stress levels in different situations at the workplace. Interviews conducted with participants were promising and indicated that workers found it easier to identify stressful factors and support them to make sense from the situations.

Another commonly used technique mentioned by \citeN{Fleck:2010cs} is the use of technology to 'see more'. This includes technologies addressing the collection of data, we could not gather without technological support, like step count data or bio-signal data. Several project have looked into supporting self-reflection; \citeN{sanches2010mind} developed a mobile application called \emph{Mind the body} which is focussed on mental health and encourages the user to reflect upon both negative and positive aspects of their behaviour. Sensing data on skin conductance and heart rate are used to determine the stress levels of the user. The mobile phone app presents the stress levels in real time. Based on the user feedback during their study, they also offer a history to view past stress levels and support reflection. \cite{Fleck:2010cs} also highlighted the important of audio and visual recording to relive our experience. \citeN{Stahl:2009fm} created \textit{Affective Diary} which is a digital diary. The mobile application has access to recorded stress data from sensors, mobile usage data, and photos. They present this data in a timeline which contains photos and arousal levels represented by shapes. They found that some users appreciated the application as it helped their self-understanding. Others experienced discomfort because the data shown highlighted bad moments. 

\subsection{Self-efficacy}
\label{self-efficacy}

Self-efficacy is a person's belief in their ability to succeed within a specific situation. \citeN{bandura1977self} stated that these beliefs are great drivers of how people think, behave and feel. A person with a strong sense of self-efficacy forms a strong sense of commitment to tasks and likes to master challenges. A person with a weak sense of self-efficacy would avoid challenging tasks altogether. In order to sustain long-term behavioural change, an individual's self-efficacy needs to remain high. There are different techniques to promote physical activity; Supporting vicarious experience from observing others being successful, giving feedback, providing participants with goals set by an interventionist and tailoring of interventions have shown to be successful strategies to increase self-efficacy for physical activity\cite{Consolvo:2009el}. In section \ref{social-influence}, we already presented a body of research on social influences and how the observations of others can support the own belief in being able to succeed.

The most effective method for maintaining self-efficacy is mastery experiences. If an individual has success and an application makes them more aware of this, it can have a positive effect on their self-confidence. Wearable applications can give rewards when users have achieved their goals. Using data in a way to set realistic goals provides the most effective methods of support for the individual. Using unrealistic and generic goals within a system could set the user on a path where they are likely to not succeed. This can cause more damage than good. The \textit{GymSkill} mobile application by \citeN{Kranz:2013ho} involves sensor data logging and activity recognition. The application works for an individual while they are balance board training. They use this to present the user with goals suitable for their ability. In the study, users liked the personalised feedback and suggestions. This shows that there is potential for this type of system to support long-term behaviour change. 

\subsection{Social Support}
\label{social-support}

Social support has been shown to have beneficial effects on our health~\cite{HoltLunstad:2015vq}; it can be defined as the "verbal and nonverbal communication between recipients and providers that reduces uncertainty about the situation, the self, the other, or the relationship, and functions to enhance a perception of personal control in one's experience" \cite[p. 19]{Albrecht:1987vp}; and it can play an important part in supporting behaviour change. \citeN{Albrecht:1987vp} describe the key factors of social support as: enhancing control, communication, and reduction of uncertainty. Different types of social support can be derived~\cite{Schaefer:1981bk}.
Emotional support is the offering of support in the form of concern, affection and caring from others. This type of support makes the individual feel valued; and it can also promote the feeling of the carried out behaviour being meaningful. Providing emotional support through a wearable interface may be positive in supporting behaviour change. It could be effective because it enhances the support that we get from our friends and family in everyday life. 
In their exploratory study - CalmMeNow - \citeN{Paredes:2011be} looked at different interventions for people in stressful situations; haptic feedback (guided breathing and acupressure), games and emotional, social support in form of text messages from loved ones. Although the results show no significant difference between these intervention types on the relaxing effect, this findings could be personality type specific. 

Another type of social support is informational support. This involves offering advice and suggestions to someone to help them solve problems. Health and fitness wearables can utilise informational support through the use of virtual trainers. \citeN{freyne2012mobile} created an application which comprises of a weight management mentor that supports dietary changes. The application would take in data about what the user was eating. It would then analyse the collected data and make suggestions for changes to the user. They created two versions of the application; the full version offered suggestions and pushed these as prompts to the user, and the control version was a simple base line application that did not have the pushing feature. The study found that users who had suggestions in the form of prompts sent to them, lost more weight than those who were not offered suggestions.

Companionship support gives an individual a sense of belonging to a social network or group. This involves encouraging the presence of others in shared social activities. \citet{mueller2007jogging} created a wearable device called \emph{Jogging over a distance}. The device uses audio pace cues to allow two people to go on a run together when they were in two different geographical locations. The spatial sound lets the user know whether the other runner is jogging in front or behind them. They found that providing companionship support in this situation was very supportive. Users could find a person to run with of similar experience and at the time of day which was most convenient to them.

\citeN{polzien2007efficacy} aimed to investigate support offered through data provided by a wearable compared to support from a human being. Their study compared results of weight loss programs supported by counselling or supportive technology. The technology used was the \emph{SenseWear Pro} wristband. This provided data to the user about their total energy expenditure and sleep efficiency. The study found that the group that used technology lost more weight within the 12 week intervention period. Those who had face to face conversations with a counsellor did not lose as much. Comparing the effectiveness over an even longer period would provide a more solid insight. The study also found that a mixture of using counselling and technology was not as effective as solely using one or the other. 

\subsection{Biofeedback} 
\label{biofeedback}

Another important topic for supporting behaviour change is biofeedback. Biofeedback involves instant feedback on an individual's own biosignals, like heart rate, respiration or brain waves. Providing meaningful feedback pre-requires that the patient understands about the meaning of the signals as well as how to influence them. Appropriate visual cues and training can help the user to instantly adjust the behaviour to positively influence their biosignals; whereby these adjustments can improve physical and mental health~\cite{Frank:2010uh}. The use of wearable technology can help to bring the traditional biofeedback therapy to the home of the user. 

\citeN{MacLean:2013ez} created \emph{MoodWings}, a project which uses biofeedback for drivers. The aim of the wearable bracelet was to make the users aware of their current stress state while they are driving. When the driver was stressed, the butterfly's wings flapped faster. The feedback positively increased their task performance and driving safety, but they found showing feedback on a driver's internal state could make them feel even more stressed. \citeN{BinYu:2014bm} presented a study looking at use of biofeedback to change the ambient environment. They used Electrocardiogram sensors to observe the heart activities; the changes in heart rate were then used to control the ambient lightning in the room. The aim was to help the user relax by actively trying to control their heart rate patterns. This subtle and intuitive interface was perceived more positively than usual Graphical User Interfaces.

\section{Challenges and Opportunities for Future Work}
\label{challenges}

In the previous sections, we surveyed the current field of wearable technologies and applications with focus on health and fitness promotion. Different techniques and psychological concepts have been applied and they bring new opportunities and challenges along.  In this section, we explain the challenges wearable technology faces at the current time. 

\subsection{Encourage Self-Motivation}

Most wearable applications aim to drive behavioural change through persuasion and creating extrinsic motivation within an individual. We showed that they achieve this by using gamification, incorporating social incentives or persuasive data representation methods. We believe that there are issues with this approach of \emph{machine-made} motivation. 

Constant motivation through external rewards can lead to the effect that we expect to get rewarded all the time to stay motivated. Effects on the removal of those external incentitives remain an area for research. These external motivators can even spoil otherwise intrinsic and enjoyable task~\cite{greene1974effects} and decrease wellbeing and feelings of self-determination~\cite{Deci:1985km}. Wearable applications currently risk manipulating data beyond recognition, especially games. 

In games, the actual sensing data is often not represented in a direct way to the user on the interface but instead it may be manipulated and hidden behind \emph{game points}. In the short term, this can create motivation and a stimulus for the user to move in order to score points. But the behaviours may not be adapted by the users every day life outside of the game. Other studies have shown that the removal of the game elements lead to a decrease in usage of a system~\cite{Thom:2012de}. A similar effect could influence the long-term effects of gamified wearable applications. Further studies have to investigate long-term motivational consequences of game elements on health behaviour changes.

In section \ref{driving-behaviour-change}, we presented the Behavioural Change Model by \citeN{fogg2009behavior} which illustrates that a balance of high motivation and high ability to execute a behaviour in combination with a trigger is needed to facilitate behavioural change. 
Instead of persuading us to change, wearables should support self-motivation and raise awareness through providing a direct, positive link between changing our behaviour and the health outcomes. Technologies, which show us the consequences of unhealthy behaviour, like the \emph{FatBelt}~\cite{pels2014fatbelt}, could raise the awareness and our self-motivation to change and avoid unpleasant outcomes. Identifying these unhealthy behaviours and providing contextualised, meaningful alternatives could help to promote a better behaviour through internal triggered motivation. 

\subsection{Design to support and motivate long-term use}

Long-term retention of health and fitness wearable devices is a big issue currently~\cite{ledger2014inside}. The majority of the studies we present in this paper take place over a short time span. Studies conducted over longer time periods are scarce and hard to find. The development of future devices and future research should take this into consideration. 

Long-term studies and the comparison of multiple approaches, like gamification and social incentives, could lead to insights on sustainable support for healthier behaviour. There is also evidence, that personal traits have to be taken into account (e.g.~\cite{Karanam:2014fd}). Long-term studies could investigate these issues and help to develop a framework for the design of wearable health promotion applications which are optimised to support different personality types in the long term. 

It is important to bring together researchers, designers and engineers with different background and expertise to address technological problems, like accurate sensing and battery life, device design and aesthetics, cognitive supporting visualisations, as well as psychological, behavioural concepts. 

\subsection{Support of change from the beginning to the end} 

Behaviour change does not happen overnight and with the same stimuli throughout the process, but it rather is happening in stages. The Transtheoretical Model from \citeN{Prochaska:1982jn} describes six stages of behaviour change:

\begin{itemize}
	\item \textit{Pre-contemplation:} In this phase, the user has no intention to change in the near future. Techniques raising the consciousness about the problem can help raising the awareness and support a transition to the next stage.
	\item \textit{Contemplation:} The user became aware of the problem and has the intention to change in the near future. If this change really happens in the near future is still unclear, but support of self-evaluation can support a transition forward.
	\item \textit{Preparation:} This stage is characterised by the serious intent to action and often involves first steps towards action. Self-efficacy support and support of commitment can help towards a transition into the action phase.
	\item \textit{Action:} People in this stage have begun to change their behaviour and make efforts to keep this going. Social support, reinforcement and substitution of problem cues with healthy cues can support the establishment of a long-term behaviour and healthy habit.
	\item \textit{Maintenance:} This stage is characterised by a successful behaviour change of more than six month and is characterised by a high self-control and self-efficacy.
\end{itemize}

The iteration through this stages is not necessarily sequential and it is possible, that users fall back. Technology that helps through all the stages could help us to identify problematic behaviour and become aware of it, take the first action steps and ensure a long-term maintenance. 

\subsection{Personal but non-intrusive interfaces for data collection and analysis}

Studies show that users want wearable health devices and applications to be personalised to their needs and situation~\cite{ananthanarayan2012persuasive, gockley2006aviva, macvean2012ifitquest}. Collecting and utilising data about the user's behaviour, personality and location can give insights in their needs and situation and support personalised and meaningful feedback. 

Collecting data about the user's behaviour and health raises issues concerning privacy and practicability. How much data needs to be collected about a user in order to be able to give a reliable representation of the user's lifestyle? This includes and overview of their choices and overall health. Obtaining many different types of data can require a collection of different sensing methods. All of these sensors obtaining data may feel intrusive for the user. 
We feel there is an opportunity to study the correlation between the amount of data collected about someone and how effective it is at describing their health in general. Current systems require many data sources to built a context around an individual's health. These countless sources can include emails, sleep patterns and location. Some users may find these systems to be intrusive and may not be comfortable with giving all of this information away. These users should still be able to have access to a system that is customisable to their comfort requirements. 

A compromise needs to be found between two elements. The first is the amount of data points a system collects to analyse. The second is the user's perception of personalisation. Can a system using minimal data input sources intelligently be just as personal as another with multiple inputs? Devices should not be intrusive, they should blend into the user's environment. Devices need to selectively analyse the minimal amount of data collected efficiently. Future work in this area will lead to integrated intelligent systems. These systems will offer a personal interface, regardless of the amount of data the user wants to provide.

\subsection{Understanding not everyone wants and needs the same type of support}

Most wearable applications offer motivation or support in one defined way. But the support offered may not be the most effective method for every person that uses the application. This leads to applications being limited to the amount of people that they can support. There is an opportunity to utilise data to find out which methods are most effective in supporting behaviour in each user. This could work well in is establishing the most effective social incentives to use within an application. 

For example, one person may be motivated more by competing with their friends but another person may prefer working as a team to achieve goals~\cite{chen2014healthytogether}. Some people may experience the self-monitoring and analysis of their own data as a sufficient tool to gain insight in their health and adjust their behaviour based on that. Other users may need active support from an wearable application through rewards, interventions or similar things. But little research has looked into providing appropriate methods support. A smart and adaptive system can analyse data and learn from the user's behaviour to work out an individual's support preferences. 

\subsection{Installing and Changing Habits}

As mentioned in Subsection \ref{subs:habits}, habits are automatic behaviours which require less cognitive load. Old habits are hard to overcome, and new ones hard to establish. Willpower and self-control are key factors, and both are dependent on our arousal and stress levels~\cite{Strack:2004bz, Segerstrom:2007fo}. Technology could support us in this processes, by detecting those arousal levels and offering contextualised support in those situations. Furthermore, technology can be used to support stress management. 

\subsection{Technology Adaption}

An important factor for wearable device facilitated behaviour change interventions is the adaption of the technology by the user. The Technology Acceptance Model by \citeN{Davis:1989fa} states that tho main factors need to be considered for adaption: the perceived ease of use and the perceived usefulness. A prerequisite for users to use the health promoting technologies is their understanding and perception of the technologies ability to enhance their performance and improve their health. New applications have to clearly communicate their role in the intervention process. In the second version of the model this criteria was extended by social influences like the subjective norm of others approving of the use of such technology and an potential increase of status within a social group~\cite{Venkatesh:2000hq}. The perceived ease of use is influenced by factors like computer self-efficacy and anxiety, the perceived controls and external resources to support the technology usage, and the personal ability for computer playfulness to try out new interactions~\cite{Venkatesh:2008ef}.  

These factors have to be considered through the design processes of wearable device user interfaces have to consider these factors to ease the use of the technology, support long-term retention and support the behaviour change.

\section{Discussions and Conclusion}
\label{conclusion}

In this paper, we reviewed the current research on utilising wearable technology to influence human health behaviour. We focused specifically on methods of data collection, manipulation and representation in wearable ecosystems. As previous studies have shown, wearable applications and the data have the power to drive positive behaviour change within an individual. By utilising methods such as gamification and social interaction, motivation can be created. This motivation increases the possibility of someone changing their health behaviours for the better. But we have found issues with using wearable sensing data as a behavioural driver. Although studies have shown it to be effective in the short term, there are issues regarding data losing its meaning to the user over time. As a response to this, it has been suggested that data and data representations should act as a facilitator for behaviour change. This can be archived by encouraging reflection and presenting the health data to accommodate cognitive theories and support the natural behavioural change process. Using data as a facilitator is showing positive hope for the development of further health wearables, but we believe that even more research is needed.

Through outlining previous studies, we believe that there are many opportunities for further research. Personalisation is an area in which more research would be beneficial. A system that can adapt to the user and recognise their needs could help to form a long-term relationship between a user and their health data. Data meaningfulness needs to remain high to ensure long-term retention between the user and their device. We suggest ways that this could be done through non-invasive collection and intelligent interpretation of health data in a way to encourage self-motivation. Wearable systems need to offer a number of different data manipulation and presentation methods. The methods would then be chosen to reflect which process the system determined to be the most effective. Ideally, research needs to be conducted that can inform the design process of future wearable technology. Ensuring long-term retention needs to be considered from the very beginning of the development process to create effective systems.

\bibliographystyle{ACM-Reference-Format-Journals}
\bibliography{biblo}

\elecappendix

\medskip

Abbreviations of Behaviour Change Theories (BCT):

\begin{itemize}
	\item SDT - Self Determination Theory \cite{Deci:1985km}
	\item SCT - Social Cognitive Theory \cite{bandura1977self}
	\item TBP - Theory of Planned Behaviour \cite{ajzen1991theory}
	\item TTM - Transtheoretical model \cite{Prochaska:1982jn}
	\item FBC - Fogg's Behaviour Change Model \cite{fogg2009behavior}
	\item FP - Fogg Persuasive \cite{fogg2002persuasive}
	\item SET - Self Efficacy Theory \cite{bandura1977self}
\end{itemize}

\begin{landscape}

\setlength{\extrarowheight}{0pt}
	\tiny
\begin{longtable}{
	p{0.1\linewidth-2\tabcolsep-1.5\fboxrule} 
	p{0.17\linewidth-2\tabcolsep-1.5\fboxrule} 
	p{0.17\linewidth -2\tabcolsep -1.5\fboxrule} 
	p{0.16\linewidth-2\tabcolsep-1.5\fboxrule} 
	p{0.07\linewidth -2\tabcolsep-1.5\fboxrule} 
	p{0.09\linewidth-2\tabcolsep-1.5\fboxrule} 
	p{0.17\linewidth-2\tabcolsep-1.5\fboxrule} 
	p{0.05\linewidth-2\tabcolsep-1.5\fboxrule}} 
\hline
Paper & Summary & Sensors and Feedback & Feedback & N & Duration & Results & BCT\\\hline\hline
\endhead
\multicolumn{7}{c}{Motivation through Gamification} 
\\\hline
\cite{bleecker2007mobzombies}
	&
		Game utilising movements from wearable in a game where avatar runs from zombies and player gets points
	&
		Accelerometer, GPS for movement data and position
	&
		game points
	&
		-
	&
		-
	&
		-
	&
		
\\\hline
\cite{Payton:2011fp}, \cite{Doran:2010va}
	&
		World of Workout exergame to motivate physical activity with quests that the user must complete by beating step count goals set up either by them or the app. 
	&
		iPhone shake sensor
	&
		new quests on a virtual quest map
	&
		10 students
	&
		2 quests
	&
		users liked the game and found it fun
	&
		
\\\hline
\cite{Stanley:2014gq}
	&
		Pervasive Accumulated Context Exergame (PACE) which passively collects activity during the day, and rewards the participant in a later sedentary computer game 
	&
		phone accelerometer, location (if on campus by looking at wifi SSID), bluetooth proximity
	&
		received game advantages in computer game, and alerts during the day 
	&
		24 players (two rounds of 12 players each)
	&
		9 days
	&
		
	&
		
\\\hline
\cite{clawson2010dancing}
	&
		Mobile and wearable health game which uses 2 wireless accelerometers worn around the users ankles as input into a social dancing game played by groups of users at the same time. 49 people were satisfied with the game but there was a lot of users finding the sensors difficult to use or that they weren’t agreeing with what they were doing.
	&
		Accelerometer for each leg
	&
		dancing game with encouraging messages
	&
		50
	&
		2 songs
	&
		found system fun and challenging, users were satisfied with experience, 
	&
		 
\\\hline
\cite{ali2006fitster}
	&
		Game with an online dashboard showing physical activity of the group members and offer capability to challenge other members of the group. 
	&
		phone accelerometer for step data
	&
		online leaderboard
	&
		-
	&
		-
	&
		-
	&
		
\\\hline
\cite{ahtinen2010let}
	&
		Designed and modelled a mobile phone game that looked at using social and play aspects to encourage physical activity called 'into'. The game works on the analogy of the user going on 'virtual trips' using their distance traveled to win rewards. People can combine and work together. 
	&
		phone accelerometer for step data
	&
		challenges within the group, presentation of the walked distance on a map
	&
		37 in groups of 2-6
	&
		1 week
	&
		users liked the app, found it understandable, and appreciated the similarity of game world
	&
		
\\\hline
\cite{chen2014healthytogether}
	&
		Developed a game to study and observe how users interact in different group gamification settings - competition, cooperation, or hybrid. App included a messaging service to allow a pair of people to talk to each other and help or taunt each other.
	&
		Fitbit pedometer
	&
		mobile dashboard
	&
		36 in pairs  (collaboration, competition, hybrid groups)
	&
		2 day warm-up, 1 week control 1 week experimental session
	&
		significant increase of activity in all groups, cooperation and hybrid effective in motivation more activity, competition lead to a negative correlation between increase of step count between pair members
	&
		
\\\hline
\cite{chuah2012wifitreasurehunt}
	&
		A location based alternate reality game that encourages users to stay physically active. The app includes ways for the user to go on tours to find hidden rewards on a real world map and also on group tours and share to social networks.
	&
		phone WiFi module to log into WiFi access points
	&
		mobile phone all with a map, access points and progress
	&
		-
	&
		-
	&
		-
	&
		
\\\hline
\cite{macvean2012ifitquest}
	&
		iFitQuest is a mobile location-aware, alternate reality exergame using google maps. Made up of mini games like ‘collect the coins’ and ‘escape the ghost’ where the user must physically move to avoid or collect things shown on screen.
	&
		mobile phone GPS and compass for locationing and navigation
	&
		Escape the Ghost: Map showing a virtual ghost avatar the players must escape by moving in the physical world; Collect the Coins: The users must collect coins shown on the virtual map while avoiding the ghosts.
	&
		25
	&
		30 minutes
	&
		users found both games enjoyable, boys enjoyed the game significantly more than girls
	&
		
\\\hline		 
\cite{gorgu2012freegaming}
	&
		Freegaming is a location-based augmented exergame. The player is navigated on a virtual map by following augmented reality clues. The status off each playerrs progress can be seen on a map in the mobile phone app. 
	&
		GPS for locationing; camea for augmented reality clues
	&
		augmented reality clues, a game map with the status of each player
	&
		-
	&
		-
	&
		-
	&
		
\\\hline
\cite{Zuckerman:2014fl}
	&
		StepByStep mobile application to promote regular walking. Quantified version with continuous measurement, daily goals and feedback on daily progression for self-motivated reflection. Gamified version with virtual rewards and social comparison.
	&
		accelerometer in phone for step count
	&
		1st study mobile app showed active minutes and progress towards goal
	&
		40
	&
		2 weeks
	&
		 significant increase of walking minutes; app raised the awareness for own activity
	&
		
\\\hline
\cite{Zuckerman:2014fl}
	&
		StepByStep mobile application to promote regular walking. Quantified version with continuous measurement, daily goals and feedback on daily progression for self-motivated reflection. Gamified version with virtual rewards and social comparison.
	&
		accelerometer in phone for step count
	&
		2nd study mobile app showed active minutes, progress towards goal and gamification elements of points and a leaderboard (in the leaderboard version)
	&
		59
	&
		10 days
	&
		 significant correlation between walking goal and active minutes for QS and points version, but not for leaderboards
	&
		
\\\hline

\multicolumn{7}{c}{Social Influence } \\\hline

\cite{buttussi2008mopet}
	&
		Generated a mobile personal trainer (MOPET), the MOPET takes in real time data from sensors and knowledge from professional trainers to provide motivation and health and safety advice. · To interact with the user there is a 3d embodied agent that can talk
	&
		GPS for locationing, heart rate monitor with a 3D accelerometer
	&
		exercise recommendations from provessional
	&
		-
	&
		-
	&
		 -
	&
		
\\\hline
\cite{lin2006fish}
	&
		Fish’n’Steps is a social computer game which links the players daily activity count to the growth of animated fish characters. 
	&
		Accelerometer for step count data
	&
		animated fish character in a bowl in the phone app
	&
		19
	&
		
	&
		 Fish‘n’Steps study indicates that participants either rose in the levels of the transtheoretical model or increased the number of daily steps
	&
		
\\\hline
\cite{anderson2007shakra}
	&
		Looks at using a mobile phone as a health promotion tool. The app tracks the daily exercise levels of users by analysing their movement. This data is then shared amongst the users group of friends. A short study found that this sharing of data encouraged the user to reflect upon the data more.
	&
		GSM cell signal strength to detect cells and movement
	&
		app shows current progress and peers progress towards goal
	&
		9
	&
		10 days
	&
		 application was perceived well by participants, no study on effect of behaviour change
	&
		
\\\hline
\cite{Cercos:2013fk}
	&
		fight sedentary through social play and collective awareness, team data shown on semi-public display, they utilise power of social relationships to change behaviour, promote shared reflective view of players, a fictional player (10k guy) to promote shared goal of 10.000 steps, utilises social comparison and showed discrepancy between goal and actual performance
	&
		Fitbit: accelerometer and altimeter
	&
		semi-public display with visualisation of step data of all participants, 2D line graph is shown
	&
		15
	&
		8 weeks
	&
		 observational results of preliminary study: when display was hanging, more and more people were interested in participating, also new conversation spaces through discussions
	&
		SDT, SCT, TTM
\\\hline
\cite{lim2011pediluma}
	&
		Pediluma is a wearable device that is worn on the user's foot, the more steps a user takes, the more the device lights up. A study was conducted and found that the use of the device did encurage an increase in physical activity but there are issues around discreetness of the device.
	&
		shoe accessory with pedometer 
	&
		LED in accessory lights up when wearer moves, dims when wearer is stationary
	&
		18
	&
		2 weeks (1 week detecting baseline, 1 week with wearable feedback)
	&
		 the device was able to increase the step count and physical activity, but people felt not too comfortable around strangers seeing the device
	&
		FP
\\\hline
\cite{Foster:2010bd}
	&
		StepMatron is a Facebook application to provide social and competitive environment to increase physical activity at the workplace 
	&
		Pedometer
	&
		non-social version: participant can see own step data in online dashboard, social: participant can additionally see step data from group members and can comment on them.
	&
		10 nurses
	&
		3w
	&
		significant increased step count in social version compared to non-social 
	&
		
\\\hline
\cite{toscos2006chick}
	&
		Chick Clique is a Mobile phone app that allows girls to talk track their step count as a group and talk about their progress. Social element worked and encouraged girls to talk to each other about health, something they would not normally do.
	&
		accelerometer as pedometer
	&
		app with group activity overview
	&
		2 groups of friends (one 4 (15-17), the other 3 (13 years))
	&
		
	&
		 comparison just tracker - tracker and app for group awareness: group one performed better with app, group two without (no significance)
	&
		
\\\hline
\cite{mauriello2014social}
	&
		Created a set of wearable devices to support group fitness, which displayed important run data to every group member on e int displays located on the back of the runners t-shirts. 
	&
		accelerometer (determine pace and distance) and heart rate
	&
		e-ink screen on back of the shirt show progress to other runners
	&
		52
	&
		1 run
	&
		display motivated to perform better; displays improve awareness of individual and group performance, helps groups stay together, and improves in-situ motivation
	&
		
\\\hline
\cite{lu2014reducing}
	&
		Designed an app called 'UOIFit' aimed at increasing physical activity levels in adolescents. The app incorporated social aspects into its design such as friending, sharing progress and collaboratively exercising with friends either in person or remotely, the fitfeed tab of the app displaying all this data. Studies were conducted with the app and found social aspects to have a positive impact on the amount of activity the user did and their BMI.
	&
		accelerometer in phone for tracking exercises
	&
		app shows fitness activity and offers social activity functions
	&
		35
	&
		6 weeks
	&
		 all participants reduced their BMI to a healthier range, strong correlation between use of social activity features ad BMI reduction
	&
		
\\\hline
\cite{Lane:2014iu}
	&
		Created a set of wearable devices to support group fitness, which displayed important run data to every group member on e int displays located on the back of the runners t-shirts. 
	&
		phone’s accelerometer and microphone to detect sleep patterns and quality, physical activity, and social activity to calculate wellbeing score
	&
		baseline group: web-dashboard showing wellbeing score: multidimensional-group: ambient phone display with fish swimming in a bowl
	&
		27
	&
		19 days
	&
		liked the ambient display, positive behaviour change, but study short 
	&
		
\\\hline
\cite{burns2012activmon}
	&
		Developed a low-complexity low- engagement interface to motivate physical activity. Activimon is a wrist worn device that has a light display that shows the user when other members of their group are being active by flashing. In studies, users were divided in their opinion about the device, some liked it but others felt it didnt provide them with enough information.
	&
		Movement of users arm, step count
	&
		wristband shows light flashing when peers are active
	&
		5
	&
		2 weeks
	&
		 usability study
	&
		
\\\hline

\multicolumn{7}{c}{Context-Aware Interfaces and Feedback} \\\hline

\cite{Oliver:2006df},
\cite{deOliveira:2008gm}
	&
		MPTrain app plays adaptive music to the runners speed, TrippleBeat app influences music based on exercise performance, advancement over MPTrain: TippleBeat considers optimal training zones, virtual competition with others, motivation through scores, glance-able interface showing 
	&
		ecg, accelerometer
	&
		MPTrain: adaptive changes in music, TrippleBeat: music changes, glancable phone screen with information
	&
		10 runners
	&
		-
	&
		compared MPTrain and TrippleBeat: increased time in trainings zone with Tripple Beat (57.1\% vs 82.8\%), all participants spend more time in optimal trainings zone, competition was valued by users, participants clearly preferred trippleBeat
	&
		
\\\hline
\cite{carroll2013food}
	&
		Looks at modifying the behaviour that people have with regards to emotional eating. Users used a mobile phone application to track emotions and to receive interventions - emotree. This helped them to find the emotions most felt when eating occurred. Then made a bra that could sense these emotions.
	&
		electrocardiogram sensor, electrodermal activity sensor, gyroscope, accelerometer, user's food and mood input to detect emotional eating patterns
	&
		
	&
		-
	&
		-
	&
		 3 studies: 2 on interventions for snacking; 1 on feasibility of wearable emotion recognition
	&
		SCT
\\\hline
\cite{Kranz:2013ho}
	&
		Looked into the feasibility of using a smart as a personalised fitness trainer. The 'GymSkill' app involves exercise descriptions, sensor data logging, activity recognition and on-top skill assessment to present data as valuable as that of a personal trainer, tailored to the users ability. The gym skill app was specific to balance board training, as the user was on the balance board, they would simply place their Andriod smart phone in the middle of the board. Testing of the system showed its potential for ensuring long-term retention in this type of application. People particularly liked the personalised feedback and suggestions more than other features.
	&
		mobile phone accelerometer and gyroscope 
	&
		Personalised feedback on exercises and suggestions on the phone
	&
		6
	&
		5 days
	&
		people especially liked the personal feedback and exercise suggestions.
	&
		SCT
\\\hline
\cite{fortmann2014waterjewel}
	&
		Waterjewel is a device to encourage its wearer to drink more water daily. It has a light up display which indicates how much of their daily goal the user has achieved but also acts as a reminder every two hours to the user that they should drink more. LEDs in bracelet would light up red or green according to users drinking behaviour. Made using Arduino lily pad and linked to an android app called 'Carbodroid'. Studies showed that people drank more when wearing the bracelet than not. So the bracelet was successful in promoting good drinking behaviour.
	&
		manual input in phone app
	&
		progress is visualised as LEDs on the wristband, progress also shown in mobile app
	&
		6
	&
		4 weeks (2 weeks in each condition)
	&
		participants drank more and more regularly with the wearable compared to the mobile app alone
	&
		
\\\hline
\cite{pels2014fatbelt}
	&
		The Fat belt is a wearable device that uses physical feedback through inflating around the stomach as a response to calorie overconsumption, simulating the long-term weight-gain associated with over-eating – isomorphism.
	&
		input of calories in mobile app
	&
		inflating belt mimicking weight gain in the stomach area
	&
		12
	&
		2 days
	&
		significant decrease in consumption over a baseline period of the same length. Seen as an extension of the users own body – gave the wearable more emotional power over the user.
	&
		
\\\hline
\cite{rajanna2014step}
	&
		A context aware health assistant system – A mobile application that encourages the user to adopt a healthy life style by performing simple and contextually suitable physical exercises. The mobile app promotes brief physical exercise after prolonged periods of inactivity by sending ‘nudges’ to the user.
	&
		accelerometer and GPS to determine activity
	&
		Nudges sent to user through their smart phone to remind them to be more active. Can be vibrations, visual or auditory.
	&
		-
	&
		-
	&
		-
	&
		FBC
\\\hline
\cite{Lin:2011cg}
	&
		Developed a context aware recommendation system called 'Motivate', which takes into consideration an individuals location and other features to offer personalised advice. The user downloads the motivate app onto their smartphone and sets up a profile for the app to base its advice around. The advice has constraints such as weather and time and models its response around these. the user then tells the app whether they intend take the advice.
	&
		GPS or GSM localisation; further information from weather services, 
	&
		App interface with recommendations for activities based on weather, location and personal preferences
	&
		6
	&
		5 weeks
	&
		Studys found the reception of this app to be mixed, with only 50\% of people replying yes to advice given to them within the app.
	&
		
\\\hline

\multicolumn{7}{c}{Support for Self-Monitoring and Reflection} \\\hline

\cite{Lin:2012jz}
	&
		mobile phone app BeWell+ to monitor sleep, physical activity and social activity to generate wellbeing score, which is presented to users. Compared two versions in their study: baseline with web dashboard and version with additional ambient display. Social feature allows to compare own wellbeing score to peers and identify role models
	&
		phone’s accelerometer and microphone
	&
		baseline group: web-dashboard: multidimensional-group: ambient phone display with fish swimming in a bowl
	&
		27
	&
		19 days
	&
		liked the ambient display, positive behaviour change, but study short 
	&
		
\\\hline
\cite{consolvo2008activity}
	&
		UbiFit is an application that uses on-body sensing and machine learning to infer people’s activities, using a personal, mobile display to encourage physical activity. The display uses the metaphor of a garden that blooms throughout the week as the user performs physical activities. 	
	&
		accelerometer and barometer to infer physical activity
	&
		glancable screen on phone with a blooming garden depending on activity levels
	&
		12
	&
		21-25 days
	&
		The technology worked reasonably well within the field study, recognising most activities correctly. Participants mentioned that the garden was motivating, often surprisingly so – worked as a constant representation of their data. For others it helped them focus on planning or simply finding time for physical activity.
	&
		TTM
\\\hline
\cite{Oliver:2006df}
	&
		Health gear consists of physiological sensors wirelessly connected to a mobile phone via Bluetooth. The data from these sensors are then manipulated and displayed to the user in a relevant way. There was an 100\% success rate in recognising cases of sleep apnea. Issues like security and privacy need to be addressed.
	&
		wearable blood oximeter for heart rate and blood oxygen levels
	&
		shows heart rate and oxygen on phone. 
	&
		20
	&
		1 night
	&
		shows heart rate and oxygen on phone. 
	&
		
\\\hline
\cite{How:2013et}
	&
		Android app to help people in rehabilitation (eg. after stroke) to improve their step symmetry. The app has three modes where the user can train, capture regular walks and compare his improvements in a history. Data can be shared with therapist via email.
	&
		Android phone accelerometer to detect step pattern
	&
		Dashboard in app with scores for symmetry of walk and stats like tome and rating of the walk
	&
		-
	&
		-
	&
		-
	&
		
\\\hline
\cite{Nerino:2013fd}
	&
		A wireless body sensor network for monitoring the exercises of rehabilitation patients of knee surgery. This supports unassisted rehabilitation of motor functions.
	&
		accelerometer, gyroscope to determine leg position and movement
	&
		GUI on a tablet for patients with feedback on correct positioning during exercises 
	&
		-
	&
		-
	&
		-
	&
		
\\\hline
\cite{zhang2013see}
	&
		Created a system that contains a wearable UV sensor and a pair of AR glasses. The glasses would change the colour of the users skin to make it look more red dependant on the amount of time the user spends out in the UV rays.
	&
		UV sensor
	&
		Augmented reality glasses showing a change in skin colour after too much sun exposure to simulate sunburn
	&
		6
	&
		-
	&
		Studies conducted found the visualisation to have a positive effect on users.
	&
		
\\\hline
\cite{Madan:2010fg}
	&
		They used mobile phone social sensing and self-assessments to identify correlations between mobile data and illness symptoms.
	&
		Bluetooth proximity to other phone users, WLAN for rough location, Call \& SMS records, daily symptoms self-assessment
	&
		no feedback
	&
		70 residents of a dormitory  
	&
		-
	&
		it is possible to determine the health status of individuals
using information gathered by mobile phones alone,
without having actual health measurements about the subject
	&
		
\\\hline
\cite{Spelmezan:2009jy}
	&
		Custom sensor and mobile phone application (phone for hosting and computation) for learning and training of physical activities. Used snowboarding beginners as example.
	&
		force sensitive resistors in shoes, Shake sensors to measure upper body rotation, BendShort sensors to measure knee flexion, 
	&
		vibro-tactile feedback through actuators placed at different places
	&
		8
	&
		1 snowboarding session
	&
		identified movements accurately, participants perceived very well tactile instructions (87
as compared to corresponding audio instructions (97
played back over earplugs while snowboarding.
	&
		
\\\hline 
\cite{Chi:2004gp}
	&
		Study Looked at putting sensors into a martial art competition (taekwondo) to see when a significant impact had been delivered to either competitor.
	&
		Force sensors, Amount of force on competitors body protector
	&
		Displays score to user on screen
	&
		4 national champions in Taekwondo
	&
		2 hours
	&
		participants and jury agreed with the scoring provided by the system; participants gave positive feedback
	&
		
\\\hline
\cite{sanches2010mind}
	&
		Different sensors brought together with a mobile app with different visualizations to help users track their stress and reflect.
	&
		Accelerometer, GSR, ECG integrated into clothing to detect movements and stress
	&
		mobile phone app, stress history view for reflection and manual pattern detection for the user
	&
		-
	&
		-
	&
		-
	&
		
\\\hline
\cite{Kocielnik:2013dt}
	&
		Framework for measuring stress in real-life conditions continuously and unobtrusively to help the users reflect their stress states and develop relief patterns
	&
		Philips DTI-2 sensor (GSR, skin temperature, ambient temperature, lightning, accelerometer), calendar entries, self-assessments
	&
		no feedback during the study, LifelogExporer showed graphs of data at the end of study
	&
		10
	&
		4 weeks
	&
		data presented to the user after the 4 weeks, LifelogExplorer to generate overview of all data 
semi-structured interviews on the meaningfulness, usefulness, and triggering of healthier behaviour; the participants were positive, that new unobtrusive sensors for long-term data measurement can help user to get feedback of their stress levels in real work environments
	&
		
\\\hline
\cite{Stahl:2009fm}
	&
		A digital diary for user-written notes as well as body sensor data and mobile phone media data to help people reflect their emotions. The tablet app showed a timeline with photos and information on the current affective state, the presence of others, and phone activity like calls and SMS. 
	&
		Bodymedia biosensor collecting skin conductance data, mobile phone photos, mobile phone usage data (SMS, calls), bluetooth proximity to detect presence of others
	&
		a tablet app allows the presentation to the collected data in a timeline. It shows photos and symbolised data on detected bluetooth device, phone calls and SMS, and figures with colour codes for current emotional state 
	&
		4
	&
		2 to 4 weeks
	&
		Their qualitative study showed that participants used the diary in very different ways to interpret and make sense of the data. They also concluded, that the measurements did not always represent the experienced feelings.  
	&
		
\\\hline
\cite{Reitberger:2014gs}
	&
		Nutriflect system with a ambient display in kitchen, shows information on healthiness of bought food
	&
		camera to scan EAN or NFC to identify foods
	&
		ambient display in form of tablet showing family process of health eating
	&
		21
	&
		2 weeks pre, 4 weeks study, 
	&
		participants liked it and rated it positively
	&
		
\\\hline

\multicolumn{7}{c}{Social Support} 
\\\hline

\cite{Paredes:2011be}
	&
		They measured stress levels of participants and offered different type of intervention to help participants relax: games, guided breathing with haptic feedback and social support from loved ones
	&
		Heart Rate sensor and GSR sensor to detect stress
	&
		3 intervention types: gaming; haptic feedback for guided breathing; emotional, social support
	&
		20
	&
		1 experimental setup
	&
		no significant difference between interventions
	&
		
\\\hline
\cite{polzien2007efficacy}
	&
		Studied whether using technology with personal counselling or using purely technology will result in a bigger weight loss. Used the Sensewear pro armband was used to determine the energy expenditure. The group that used solely technology lost the most weight, followed by the people that used solely counselling.
	&
		Sensewear pro armband for determining energy expenditure, self-monitored calorie intake
	&
		Inteructions on calorie intake and exercise regime, the technology group also had access to their energy expenditure data from Sensewear
	&
		58
	&
		12 weeks
	&
		participants using the Sensewear lost ~2kg more weight than group without technology support (not significant)
	&
		SET
\\\hline

\end{longtable}

\end{landscape}

\end{document}